%% file: main.tex
\documentclass[11pt, a4paper, twocolumn, internal, copyright]{louvresae}

\pdftrailerid{redacted}

\makeatletter
\renewcommand\bibentry[1]{\nocite{#1}{\frenchspacing\@nameuse{BR@r@#1\@extra@b@citeb}}}
\makeatother

\usepackage{kantlipsum, lipsum}
\usepackage{dsfont}
\usepackage{louvresae-colors}
\usepackage{listings}%
\usepackage{placeins}
\usepackage{subcaption}
\usepackage{float}
\usepackage{inconsolata}
\usepackage[hang,flushmargin]{footmisc}

\renewcommand{\internalonly}{}
\renewcommand{\today}{6 April 2026}
\renewcommand{\copyrightext}{}

\usepackage[authoryear, sort&compress, round]{natbib}

\graphicspath{{figures/}}

\title{The Impact of AI-Generated \\ Text on the Internet}
\newcommand{\shorttitle}{The Impact of AI-Generated Text on the Internet}

\correspondingauthor{maty@stanford.edu}
\codeanddataurl{https://ai-on-the-internet.github.io}

\keywords{Artificial Intelligence, Large Language Models, AI-Generated Text, Internet, Online Discourse}
\paperurl{} % arxiv.org/abs/tbd

\reportnumber{} % Leave blank if n/a

\author[1]{Jonas Dolezal}
\author[2]{Sawood Alam}
\author[2]{Mark Graham}
\author[3]{Maty Bohacek}

\affil[1]{Imperial College London}
\affil[2]{Internet Archive}
\affil[3]{Stanford University}

\begin{document}

\input{sec/0_abstract}

\maketitle

% Include paper content from external files

\input{sec/1_intro}

\input{sec/2_results}

\input{sec/3_discussion}

\input{sec/4_methods}

\section*{Acknowledgements}
The authors thank the Internet Archive for providing the Wayback Machine data and their extraordinary support throughout this project and Pangram for providing a research grant supporting the use of their API. We also thank Liam Dugan, Hany Farid, Daphne Ippolito, Shayne Longpre, and Alexander Wang for helpful discussions (listed in alphabetical order).

\section*{Competing Interests}
Pangram had no role in study design, data collection, analysis, interpretation, or the decision to publish. We only applied for the research grant after conducting the robustness analysis and deciding to use their detector.

% \section*{Citing this work}
% The final version of this work was published online (provide venue, date and digital object
% identifier (doi, if available). \textit{Cite as:} \citeas{dolezal2026impact}.

% Bibliography components
\bibliographystyle{abbrvnat}
\nobibliography*
\bibliography{refs}

\newpage
\onecolumn

\appendix

\input{sec/5_appendix}

% Some other useful sections you might consider having in your report.

% If you add a bibtex entry of your own paper (this paper), you can
% show its full citation inline using \citeas, as above. Note that
% this citation removes the trailing full stop. To make \citeas work,
% you need to load the bibliography data. This can be done in two
% ways:
%
%    1. If you already have a printed bibliography with \bibliography{...},
%       then add the command "\nobibliography*", no arguments, before that.
%    2. If you don't otherwise print a bibliography, add the command
%       \nobibliography{...} at the end of your document.

\end{document}

%% file: sec/0_abstract.tex
\begin{abstract}
The proliferation of AI-generated and AI-assisted text on the internet is feared to contribute to a degradation in semantic and stylistic diversity, factual accuracy, and other negative developments (sometimes subsumed under the ``Dead Internet Theory''). What has hindered answering these questions is that it has not been understood just how much of the internet is actually AI-generated or AI-edited. To this end, we construct a representative sample of websites published on the internet between 2022 and 2025 using the Internet Archive, and apply a state-of-the-art AI text detector on them. We find that by mid-2025, roughly 35\% of newly published websites were classified as AI-generated or AI-assisted, up from zero before ChatGPT's launch in late 2022. We also find statistically significant evidence for some of the identified hypotheses; for example, that increases in AI-generated text on the internet correlate negatively with semantic diversity and positively with the prevalence of positive sentiment. We do not, however, find statistically significant evidence supporting the hypothesis that an increased rate of AI-generated text on the internet decreases factual accuracy or stylistic diversity. Notably, this diverges from public perception, which we measure in a user study, where the majority of US adults turned out to believe in all four of the above-mentioned hypotheses. Individuals who do not use AI or use it infrequently tend to believe in these negative impacts more than those who use it frequently; similarly, individuals who hold negative views of AI tend to believe in these hypotheses more than those with favorable views of the technology.
\end{abstract}

%% file: sec/1_intro.tex
\noindent Ever since ChatGPT first made large language models (LLMs) available to the wider public in 2022, which was followed by mass adoption, there have been concerns about the impact of AI-generated text (as well as AI-generated content in other modalities) on the internet and online discourse~\citep{muzumdar2025dead, ferrara2026generative}. Specifically, many known limitations and failure modes of LLMs, including factual hallucinations~\citep{huang2025survey}, sycophancy~\citep{malmqvist2025sycophancy}, verbosity~\citep{saito2023verbosity}, and more, have raised concerns that unchecked proliferation of such content could reduce the overall quality of internet content~\citep{xing2025llms, shumailov2024ai}. These hypotheses are sometimes subsumed under the ``Dead Internet Theory,'' which they loosely expand, but which, on its own, predates the widespread use of LLMs~\citep{muzumdar2025dead}. These hypotheses have been difficult to verify, primarily because there is limited understanding of how much internet content is actually AI-generated~\citep{spennemann2025delving, santy2025incentives}. In this paper, we attempt to address these questions. We concern ourselves only with LLM-generated text, leaving other modalities for future work, and use LLM-generated and AI-generated interchangeably.

Verifying the above-mentioned hypotheses about AI-generated text on the internet is difficult for several reasons. First, obtaining representative samples of text from the internet is challenging~\citep{thompson2024improved}. As a result, existing analyses have mainly focused on restricted subsets of the internet, such as specific social media platforms~\citep{la2025machines, sun2025we, matatov2024examining} or news sources~\citep{russell2025ai}, scientific publishing~\citep{liang2024monitoring, kobak2025delving}, software repositories~\citep{daniotti2026using}, or translations~\citep{thompson2024shocking}. Second, detecting AI-generated content is itself a difficult problem with many known challenges~\citep{fraser2025detecting, sadasivan2025can, dawkins2025detection}. To date, AI detection for image and video content has been believed to be more accurate than for text~\citep{wu2025survey}. Thus, the few existing inquiries in this domain have predominantly focused on image and video modalities rather than text~\citep{matatov2024examining}. To the best of our knowledge, no prior study has analyzed the impact of AI-generated text on the internet as a whole. We attempt to do so here.

We pose the following research questions: (RQ1) What are the generally held beliefs about the impacts of AI-generated text on the internet among adults in the United States? (RQ2) How many websites on the internet, published between 2022 and 2025, contain AI-generated text? (RQ3) Are the hypotheses about the impact of AI-generated text on the internet from RQ1 correct? To answer RQ1, we conduct a study of a representative sample of US adults to test beliefs about six hypotheses identified by the research team through exploratory environmental scanning and thematic analysis of online discourse. We then sample representative sets of websites from the Internet Archive published between 2022 and 2025 and detect AI-generated text using the detector that performed best in our independent evaluation. Finally, we conduct a series of quantitative experiments to evaluate the hypothesized impacts against this data.

%% file: sec/2_results.tex
\section{Results}
\label{sec:results}

\begin{figure*}[t]
    \centering
    \includegraphics[width=1.0\linewidth]{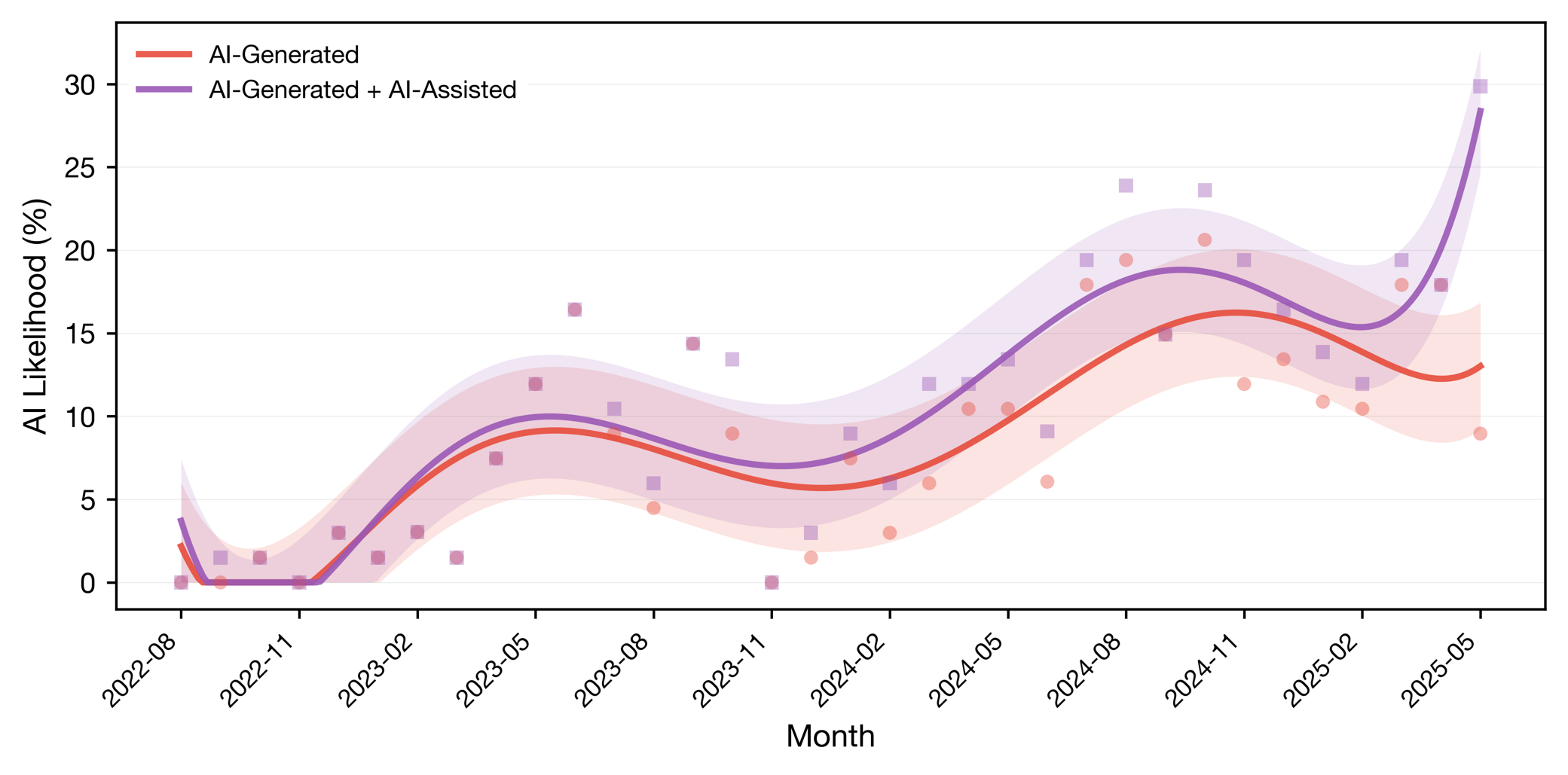}
    \caption{\textbf{AI-generated Text on the Internet from Mid-2022 to Mid-2025.} The figure shows the proportion of websites classified as fully AI-generated (red) and AI-generated or AI-assisted (purple) based on Pangram v3 detection applied to representative samples obtained from the Internet Archive. Curves represent smoothed estimates.}
    \label{fig:ai_prevalence_overall}
\end{figure*}

We found that the prevalence of AI-generated and AI-assisted websites has been growing since the launch of ChatGPT in November 2022. By the first half of 2025, as much as $35\%$ of websites uploaded to the internet in a given month were AI-generated or AI-assisted. The share of AI-generated and AI-assisted websites over time is shown in Figure~\ref{fig:ai_prevalence_overall}.

We next review each of the six hypotheses about the impact of AI-generated and AI-assisted text on the internet, presenting the results of our participant survey as well as a quantitative analysis against internet data from the Internet Archive. The aggregate AI likelihood score represents the likelihood of text to be AI-generated and AI-assisted in a given sample.

\begin{figure*}[t]
    \centering
    \begin{minipage}[t]{1.0\linewidth}
        \centering (a)
    \end{minipage}
    \includegraphics[width=1.0\linewidth]{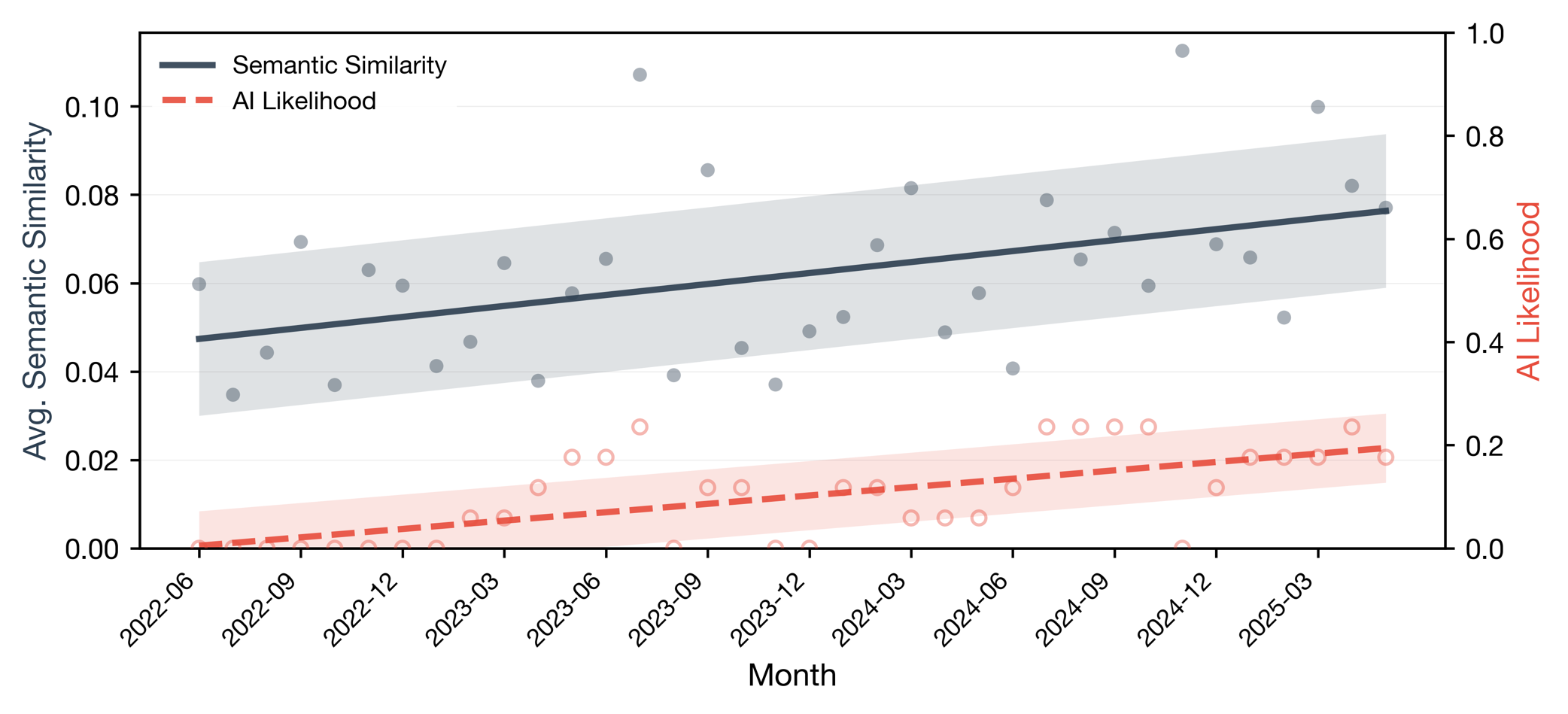}
    \vspace{1em}
    \includegraphics[width=1.0\linewidth]{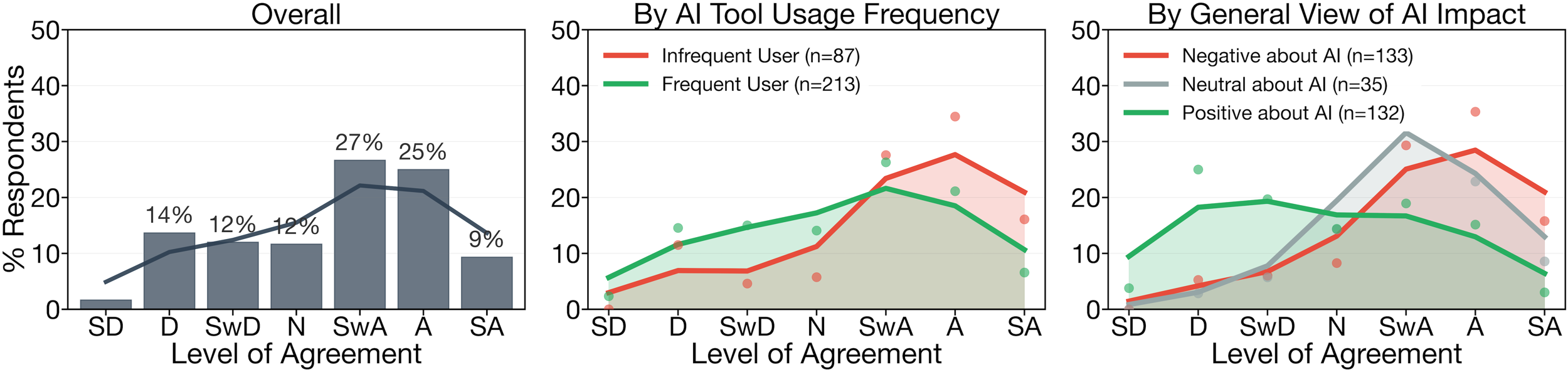}
    \begin{minipage}[t]{0.333\linewidth}
        \centering (b)
    \end{minipage}%
    \begin{minipage}[t]{0.333\linewidth}
        \centering (c)
    \end{minipage}%
    \begin{minipage}[t]{0.333\linewidth}
        \centering (d)
    \end{minipage}
    \caption{\textbf{Results for Hyp. 1: Semantic Contraction.} The figure shows results for the Semantic Contraction Hypothesis from the participant study (RQ1) and quantitative analysis of randomly sampled websites from the Internet Archive (RQ3). In (a), the average pairwise cosine similarity of semantic embeddings is plotted against AI Likelihood score, which combines the rate of AI-generated and AI-assisted samples, as detected by Pangram v3 ($\rho=0.47$, $p=0.004$). The overall results of the participant study are shown in (b), with responses ranging from Strongly Disagree (SD) to Strongly Agree (SA). These are broken down by AI usage frequency in (c) and general view of AI impact in (d).}
    \label{fig:hyp_1}
\end{figure*}

\paragraph{The Semantic Contraction Hypothesis (Hyp. 1).} The statement of this hypothesis is the following: ``As AI text becomes more common on the internet, the range of unique ideas and diverse viewpoints shrinks.'' The results are shown in Figure~\ref{fig:hyp_1}. $60.9\%$ of respondents lean towards agreement with this statement, $11.7\%$ are neutral, and $27.4\%$ lean towards disagreement. This has been translated into the following measurable signal: the average pairwise cosine similarity of semantic embeddings within a monthly sample. If the hypothesis is correct, we posit this signal would be positively correlated with the aggregate AI likelihood score. The null hypothesis ($H_0$, i.e., the correlation $\rho=0$) was rejected ($\rho=0.47$, $p=0.004$), confirming the hypothesis. Across all evaluated months, the average semantic similarity between websites predicted to be AI-generated or AI-assisted was $33\%$ higher than that of non-AI websites (semantic similarity scores of $0.0701$ vs. $0.0526$, respectively).

\paragraph{The Truth Decay Hypothesis (Hyp. 2).} The statement of this hypothesis is the following: ``As AI content becomes more common on the internet, I am encountering factually incorrect information and hallucinations more frequently.'' $75.1\%$ of respondents lean towards agreement with this statement, $14.3\%$ are neutral and $20.6\%$ lean towards disagreement. This has been translated into the following measurable signal: the average rate of factually incorrect statements within a monthly sample. If the hypothesis is correct, we posit this signal would be positively correlated with the aggregate AI likelihood score. The null hypothesis ($H_0$, i.e., the correlation $\rho=0$) was not rejected ($\rho=-0.19$, $p=0.27$).

\begin{figure*}[t]
    \centering
    \begin{minipage}[t]{1.0\linewidth}
        \centering (a)
    \end{minipage}
    \includegraphics[width=1.0\linewidth]{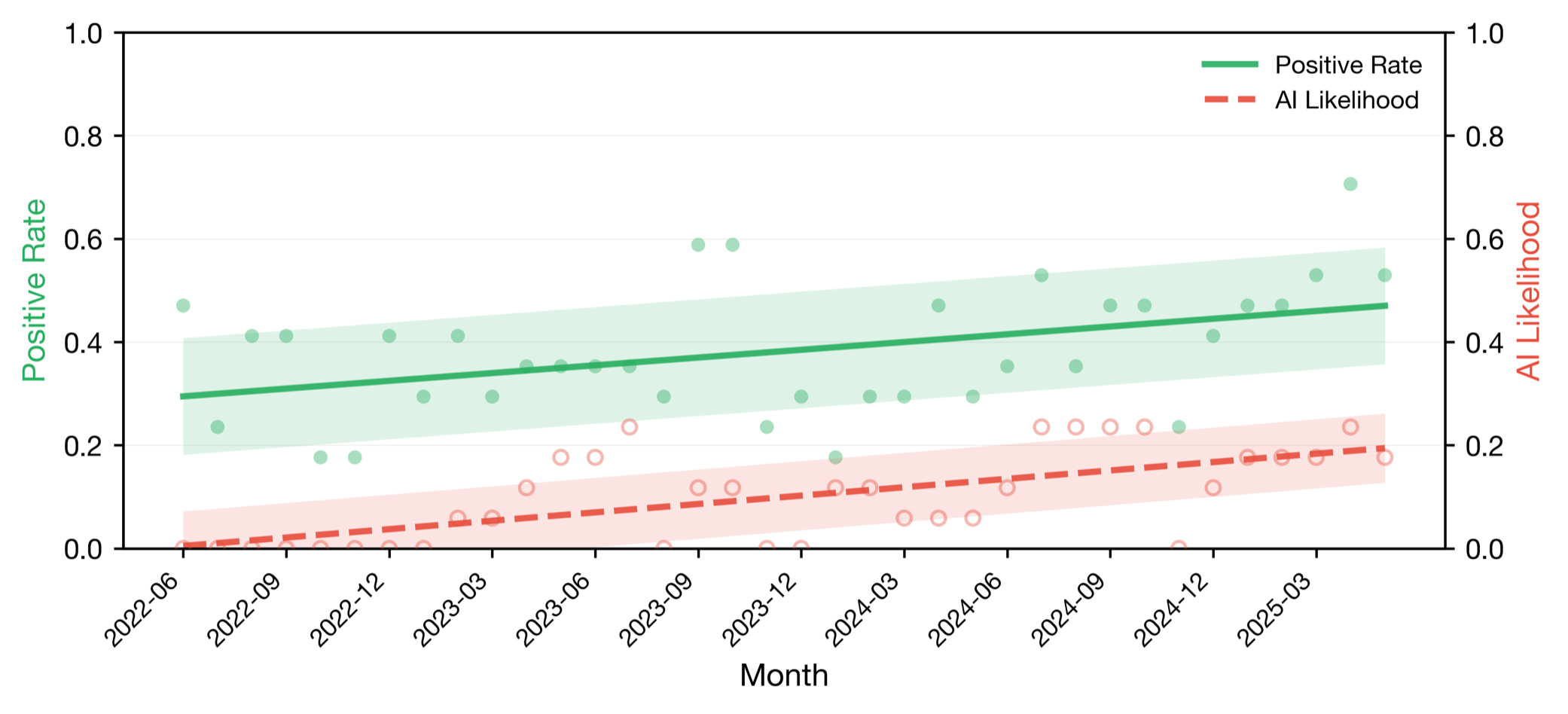}
    \vspace{1em}
    \includegraphics[width=1.0\linewidth]{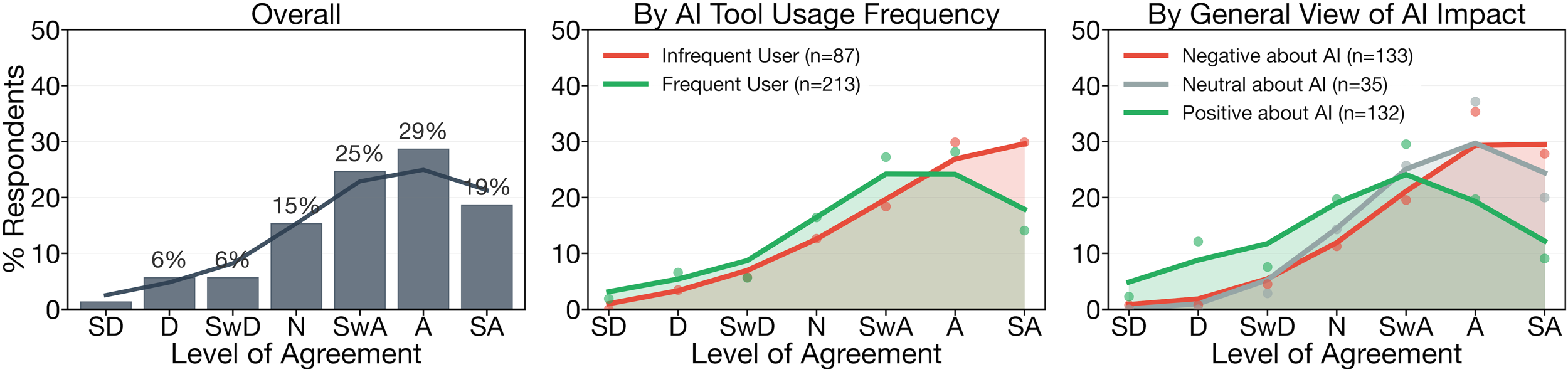}
    \begin{minipage}[t]{0.333\linewidth}
        \centering (b)
    \end{minipage}%
    \begin{minipage}[t]{0.333\linewidth}
        \centering (c)
    \end{minipage}%
    \begin{minipage}[t]{0.333\linewidth}
        \centering (d)
    \end{minipage}
    \caption{\textbf{Results for Hyp. 3: Positivity Shift.} The figure shows results for the Positivity Shift Hypothesis from the participant study (RQ1) and quantitative analysis of randomly sampled websites from the Internet Archive (RQ3). In (a), the rate of positive documents (classified by sentiment analysis) is plotted against AI Likelihood score, which combines the rate of AI-generated and AI-assisted samples, as detected by Pangram v3 ($\rho=0.56$, $p=0.0003$). The overall results of the participant study are shown in (b), with responses ranging from Strongly Disagree (SD) to Strongly Agree (SA). These are broken down by AI usage frequency in (c) and general view of AI impact in (d).}
    \label{fig:hyp_3}
\end{figure*}

\paragraph{The Positivity Shift Hypothesis (Hyp. 3).} The statement of this hypothesis is the following: ``As AI content becomes more common on the internet, online writing feels increasingly sanitized and artificially cheerful.'' The results are shown in Figure~\ref{fig:hyp_3}. $72.0\%$ of respondents lean towards agreement with this statement, $15.3\%$ are neutral and $12.7\%$ lean towards disagreement. This has been translated into the following measurable signal: the rate of positive documents (i.e., classified as positive by sentiment analysis) within a monthly sample. If the hypothesis is correct, we posit this signal would be positively correlated with the aggregate AI likelihood score. The null hypothesis ($H_0$, i.e., the correlation $\rho=0$) was rejected ($\rho=0.56$, $p=0.0003$), confirming the hypothesis. Across all evaluated months, the average positive sentiment score of AI-generated or AI-assisted was $107\%$ higher than that of non-AI websites (sentiment scores of $0.7042$ vs. $0.3400$, respectively).

\paragraph{The Epistemic Island Hypothesis (Hyp. 4).} The statement of this hypothesis is the following: ``As AI content becomes more common on the internet, articles are increasingly providing answers without including links to external sources.'' $69.9\%$ of respondents lean towards agreement with this statement, $18.7\%$ are neutral and $11.4\%$ lean towards disagreement. This has been translated into the following measurable signal: the density of outbound link tags. If the hypothesis is correct, we posit this signal would be inversely correlated with the aggregate AI likelihood score. The null hypothesis ($H_0$, i.e., the correlation $\rho=0$) was not rejected ($\rho=-0.12$, $p=0.48$).

\paragraph{The Entropy Dilution Hypothesis (Hyp. 5).} The statement of this hypothesis is the following: ``As AI content becomes more common on the internet, content is becoming significantly longer in word count while having lower semantic density.'' $60.7\%$ of respondents lean towards agreement with this statement, $13.8\%$ are neutral and $25.5\%$ lean towards disagreement. This has been translated into the following measurable signal:  the Gzip compression ratio calculated as the fraction of raw document size over compressed size. If the hypothesis is correct, we posit this signal would be positively correlated with the aggregate AI likelihood score. The null hypothesis ($H_0$, i.e., the correlation $\rho=0$) was not rejected ($\rho=-0.02$, $p=0.89$).

\paragraph{The Stylistic Monoculture Hypothesis (Hyp. 6).} The statement of this hypothesis is the following: ``As AI content becomes more common on the internet, distinct individual writing styles are disappearing in favor of a generic, uniform voice.'' $83.0\%$ of respondents lean towards agreement with this statement, $9.4\%$ are neutral and $7.6\%$ lean towards disagreement. This has been translated into the following measurable signal: the average pairwise cosine similarity of document writing style embeddings within a monthly sample. If the hypothesis is correct, we posit this signal would be positively correlated with the aggregate AI likelihood score. The null hypothesis ($H_0$, i.e., the correlation $\rho=0$) was not rejected ($\rho=0.24$, $p=0.17$).

\paragraph{Impact of AI Usage and Favorability.} For all tested hypotheses, participants who use AI infrequently were more likely to believe that the negative impact of AI-generated content on the internet is real than those who use AI tools regularly, with a pooled agreement rate of $88.3\%$ versus $76.2\%$ among non-neutral respondents (a difference of $12.1$ percentage points). Similarly, participants with a less favorable view of AI's general impact on society were more likely to believe in the negative impact of AI-generated content on the internet than those with a favorable or neutral view of the technology’s societal impact, with a pooled agreement rate of $91.3\%$ versus $71.1\%$ (a difference of $20.2$ percentage points).

%% file: sec/3_discussion.tex
\section{Discussion}
\label{sec:discussion}

Our study shows a shift in the composition of the open web, estimating that as much as 35\% of newly published websites by mid-2025 have been AI-generated or AI-assisted. Notably, we find a divergence between the impacts of this shift on online discourse and the public perception of this phenomenon. While our survey (RQ1) reveals a public concern about systemic truth decay (Hyp. 2) and stylistic homogenization (Hyp. 6) as a result of AI-generated text proliferation, our web-scale analysis (RQ3) does not yield statistically significant evidence of macro-level degradation in factual accuracy or a strict stylistic monoculture.

This divergence suggests that the immediate threat to online discourse may be of an epistemic nature rather than purely factual. As AI-generated text becomes ubiquitous and indistinguishable from human writing~\citep{jakesch2023human, porter2024ai, chein2024human}, users may discount the credibility of all online information (operating on the principle of ``reality apathy'' or, when used maliciously, the ``liar's dividend'')~\citep{liu2025seeing,altay2024people,chesney2019deep, schiff2025liar}. This could potentially alter online news consumption patterns, driving users toward more insular information ecosystems~\citep{jacob2025chat, kitchens2020understanding}. If true, infrequent users of AI tools and general skeptics of the technology would likely be most affected, since we find they harbor a deeper concern about the negative impact of AI-generated text on the internet than their counterparts.

Rather than an explosion of falsehoods, the footprint of AI proliferation on the internet manifests primarily as semantic contraction (Hyp. 1) and an artificial positivity shift (Hyp. 3). The increase in semantic similarity among AI-generated text compared to human-written text implies that the online Overton window may be narrowing, as LLMs optimize for outputs that fall within a more constrained distribution closer to the average of the training distribution~\citep{agarwal2025ai, dohmatob2024model, zhang2025verbalized}. Simultaneously, the increase in positive sentiment, symptomatic of the sycophantic and overoptimistic nature of existing LLMs~\citep{malmqvist2025sycophancy, sharma2023towards, chen2025helpfulness}, implies that the discourse may be becoming more sanitized. Pluralistic online engagement relies on friction, debate, and the processing of diverse societal realities, including the negative ones~\citep{mouffe1999deliberative, lasser2025designing}; an environment flooded with cheerful, homogenized text may marginalize human dissent~\citep{daryani2026homogenizing, oh2025does}. This semantic contraction could quietly pacify or homogenize public attitudes at a scale without employing overt disinformation~\citep{daryani2026homogenizing, agarwal2025ai}.

The widespread proliferation of AI-generated content driving this homogenization of discourse appears to be largely driven by economic incentives rather than coordinated intent~\citep{santy2025incentives, zhang2024impact}. However, this crowding out of human involvement in the production of text on the internet exposes a vulnerability in existing platform governance: while online platforms possess robust infrastructures to detect and moderate overt harms, such as hate speech~\citep{hee2024recent, gorwa2020algorithmic} or, to some extent, factual inaccuracies~\citep{westlund2024problem, tokita2024measuring}, they are unequipped to govern for semantic diversity or epistemic quality~\citep{palla2025policy, lasser2025designing, gorwa2020algorithmic}.

In addition to impacts on online discourse, this shift in the makeup of the internet carries immediate consequences for AI research and development. The 35\% prevalence of AI-generated and AI-assisted text transforms the theoretical risk of model collapse~\citep{shumailov2024ai, schaeffer2025position}, wherein future AI models degrade after recursively ingesting AI-generated data, into an empirical concern. Future foundation models trained on contemporary data crawled from the internet will inevitably ingest a dataset that is, to a large extent, AI-generated and substantially less semantically diverse. While this may have a practical impact on pre-training, post-training and alignment stages will likely not be impacted, as they mostly utilize newly generated or environment-based data~\citep{gan2024towards, dohmatob2024strong}. It also remains an open question whether these recursive degradations are triggered by self-ingestion within a single model lineage or if they persist across a heterogeneous ecosystem of disparate models~\citep{gerstgrasser2024is, alemohammad2023self}.

While our findings introduce concerns for productive democratic discourse on the internet, AI-generated text online should not be understood as inherently negative or as having intrinsic moral value. We see many scenarios in which it may democratize access to online conversation, such as empowering non-native speakers and individuals with varying literacy levels to participate more fluently in the online public sphere~\citep{baldrich2025artificial, tafazoli2024exploring, kalantzis2025literacy}. Other positive use cases might include automated summarization of complicated documents for information accessibility~\citep{marturi2025llm}, and the mass-scale localization of global knowledge for underserved, low-resource languages~\citep{ankinina2025are, merx2024low}.

Countering the unintended consequences of AI proliferation on the internet requires acknowledging the fundamental limits of post-hoc detection. While the detection of AI-generated text remains reliable for now (see Appendix~\ref{sec:appendix_robustness}), in some contexts it is intractable (e.g., when the text is very short), and the confidence of these methods may change over time. Mandates relying on retroactive detection or easily circumvented text watermarking~\citep{nemecek2025watermarking, cheng2025revealing, rijsbosch2025adoption, migliorini2024china}, which have already been passed in numerous countries~\citep{aiact, euaiframework}, may therefore be inadequate. Preserving an open discourse with verifiable human participation may instead require a pivot toward cryptographically verifying human provenance (e.g., C2PA-like standards)~\citep{c2pa_spec, c2pa_privacy} and recalibrating search and recommendation algorithms to reward semantic diversity and verified human origin over raw content volume or engagement~\citep{lasser2025designing, yu2024nudging}.

Even though factual accuracy may not be a concern in online discourse at present, future research should examine whether factual accuracy degrades as recursive training feedback loops accelerate across subsequent model generations. Additionally, while this study offers a baseline for text, analogous web-scale prevalence studies are urgently needed for multimodal content~\citep{croitoru2024deepfake, chandra2025deepfake}. In the context of global elections, researchers should carefully distinguish between the ambient proliferation of financially motivated AI content and the targeted intent of adversarial campaigns, as human intent remains an essential variable in protecting democratic discourse.

It is important to note that the impact of AI-generated imagery may be fundamentally different from text; while text-based synthetic proliferation primarily causes semantic contraction, deepfake imagery poses a more direct threat to visual evidence and can trigger more visceral forms of systemic truth decay due to the historical trust placed in photographic documentation.

%% file: sec/4_methods.tex
\section{Methods}
\label{sec:methods}

\subsection{Data Collection}
\label{sec:data_collection}

To obtain a representative sample of websites published on the internet, we use the longitudinal URL sampling methodology introduced by~\citet{garg2025longitudinal} to draw samples from the Wayback Machine at the Internet Archive. This approach employs multi-dimensional stratified sampling of the Internet Archive's CDX index, which catalogs every archived web page. Specifically, URLs are sampled along several dimensions---time of first archival, MIME type, URL depth, and top-level domain---to mitigate biases arising from the Archive's increasing crawl capacity over time and the over-representation of popular domains. Logarithmic-scale downsampling is applied to reduce the influence of highly archived domains, and top-level URLs are extracted from deep links to upsample earlier periods. The resulting sample is designed to approximate a uniform random draw from the population of publicly accessible web pages archived during the sampling period.

We sample websites from the Internet Archive spanning 33 monthly intervals from August 2022 to May 2025. For each sampled URL, we retrieve the oldest available archived snapshot via the Wayback Machine's CDX Server API~\citep{garg2025longitudinal}.\footnote{We note any modifications on top of the sampling methodology used by \citet{garg2025longitudinal} in Appendix~\ref{sec:appendix_sampling}.} The raw HTML of each snapshot is downloaded and stored locally for subsequent processing.

From each archived HTML page, we extract the visible text---i.e., the textual content that would be rendered and visible to a user visiting the page in a web browser. We accomplish this using the Trafilatura~\citep{barbaresi2021trafilatura} library, which is designed to isolate the main textual content of webpages while removing boilerplate elements such as navigation menus, headers, footers, and other non-content components. The extracted content is segmented into paragraphs, and the longest paragraph (measured by word count) is selected as the representative text sample for that page. If the resulting paragraph contains fewer than 100 words, the document is excluded from further analysis. Because interface elements injected by the Wayback Machine (e.g., replay banners or error messages) are typically short, these steps make it unlikely that such artifacts appear in the final dataset. In addition, language detection is applied using the \texttt{langdetect}~\citep{langdetectpython} library, and only English text is retained.

\subsection{Participant Study}
\label{sec:participant_study}

To assess public beliefs about the impact of AI-generated text on the internet (RQ1), we conducted a survey of adults in the United States, recruited through Prolific. The sample was stratified to be representative of the US adult population with respect to age, sex, and ethnicity.

The study was administered as an online form in three parts, each corresponding to a subset of the six hypotheses. Part~1 addressed Hypotheses~1--3 ($n=303$), Part~2 addressed Hypotheses~4 and~6 ($n=301$), and Part~3 addressed Hypothesis~5 ($n=299$). In total, $N=903$ responses were collected from $853$ unique participants. Each part presented participants with a hypothesis statement and asked them to indicate their level of agreement on a 7-point Likert scale (Strongly Disagree, Disagree, Somewhat Disagree, Neither Agree nor Disagree, Somewhat Agree, Agree, Strongly Agree). In addition, each part collected two covariates: frequency of AI tool usage (Never, Monthly, Weekly, Daily) and general view of AI's impact on society (7-point scale from Very Negative to Very Positive).

The combined sample had a mean age of $45.4$ years ($\text{SD}=15.73$, range: 18--84) and was $50.9\%$ female and $48.6\%$ male. The ethnic composition was $62.6\%$ White, $12.0\%$ Black, $11.1\%$ Mixed, $7.6\%$ Other, and $6.3\%$ Asian. Nearly all participants ($99.6\%$) resided in the United States. Regarding AI tool usage, $39.8\%$ reported daily use, $31.3\%$ weekly, $17.1\%$ monthly, and $11.2\%$ never. In terms of general views of AI's impact on society, $46.0\%$ held a positive view (Somewhat Positive, Positive, or Very Positive), $40.7\%$ held a negative view (Somewhat Negative, Negative, or Very Negative), and $12.6\%$ were neutral.

\subsection{AI-Generated Text Detection}
\label{sec:ai_detection}

Detecting AI-generated text is a prerequisite for all subsequent analyses. We evaluate four detectors: Binoculars~\citep{hans2024spotting}, Desklib~\citep{desklib2025aitextdetector}, DivEye~\citep{basani2025diversity}, and the Pangram v3 commercial API~\citep{pangram_api_docs}. These were chosen based on their RAID benchmark performance~\citep{dugan2024raid} and availability. While AI text detection is known to have its limitations, we believe it provides the best estimate of AI text prevalence in in our empirical setting (i.e., a collection of texts without available watermarking). We create a custom set of tests to better understand these limitations for our use case, and choose the most suitable detector, as reported below.

We compare the four above-mentioned detectors across five dimensions: (1)~text length sensitivity, testing detection performance on texts ranging from $1$ to $500$ words; (2)~HTML robustness, comparing detection accuracy on identical AI-generated text in plain versus HTML-embedded formats; (3)~model family, evaluating detection of outputs from GPT-4o, Claude, and Gemini; (4)~model version, testing across OpenAI model versions from davinci-002 to GPT-4o; and (5)~multilingual robustness. The full results of this robustness analysis are reported in Appendix~\ref{sec:appendix_robustness}.

Based on this evaluation, we selected the Pangram v3 API for our primary analyses. Pangram v3 outperformed the three remaining methods across most robustness dimensions, achieving perfect accuracy on texts exceeding $50$ words across languages and in both raw text and HTML-wrapped formats. Binoculars and DivEye proved particularly unreliable: Binoculars exhibited an $11.4$ percentage point drop in detection rate when AI-generated text was embedded in HTML, and showed substantially higher difficulty detecting Claude-generated text; DivEye showed no overlap between plain text and HTML-embedded score distributions and failed to separate AI-generated from human-written text across most non-English languages. Desklib performed comparably to Pangram v3 on text length and model family robustness, and even outperformed it on model version robustness; however, it underperformed on HTML versus plain text and language robustness, which we consider more critical given that our corpus consists of multilingual archived web pages. In contrast, Pangram v3 maintained stable performance across all five evaluated dimensions. An additional advantage of Pangram v3 over the three other detectors is that it operates in a three-way classification scheme --- AI-generated, AI-assisted, and human-written --- that provides richer signal than binary detection.

Pangram v3 classifies each input text into three categories: (1)~fully AI-generated (\texttt{fraction\_ai}), (2)~AI-assisted, i.e., human-written with AI involvement (\texttt{fraction\_ai\_assisted}), and (3)~fully human-written (\texttt{fraction\_human}), where the three fractions sum to approximately one. The detector operates over sliding windows of the input text and aggregates segment-level predictions into document-level scores.

For each website in our sample, we report three scores corresponding to these categories. We define the \emph{aggregate AI likelihood score} for a given monthly sample as a combined metric incorporating both the fully AI-generated and AI-assisted fractions, which serves as the primary independent variable in our hypothesis tests. In the results, we additionally report the proportion of websites classified as fully AI-generated (i.e., where \texttt{fraction\_ai} exceeds a threshold) and the proportion classified as either AI-generated or AI-assisted.

\subsection{Hypothesis Verification}
\label{sec:hypothesis_verification}

We test six hypotheses about the impact of AI-generated text on the internet. For each hypothesis, we define a measurable signal, compute it for each monthly sample of websites, and test whether it correlates with the aggregate AI likelihood score across months. All correlations are evaluated using the Pearson correlation coefficient; we reject the null hypothesis ($H_0$: $\rho = 0$) at the $\alpha = 0.05$ significance level.

\paragraph{Hypothesis 1: Semantic Contraction.}
We operationalize semantic diversity as the average pairwise cosine similarity of document-level semantic embeddings within each monthly sample. Embeddings are computed using the \texttt{all-MiniLM-L6-v2} sentence embedding model~\citep{reimers2019sentence} from the \texttt{sentence-transformers} library\footnote{\url{https://huggingface.co/sentence-transformers/all-MiniLM-L6-v2}}. Each document's extracted visible text is encoded into a 384-dimensional dense vector. For samples with $n \leq 500$ documents, we compute all $\binom{n}{2}$ pairwise cosine similarities using \texttt{scipy.spatial.distance.pdist}; for larger samples, we estimate the mean pairwise similarity by randomly sampling up to 10{,}000 document pairs. Embeddings are L2-normalized prior to similarity computation. A positive correlation between this signal and the aggregate AI likelihood score would indicate semantic contraction.

\paragraph{Hypothesis 2: Truth Decay.}
We measure factual accuracy using a two-stage pipeline: automated claim extraction followed by human fact-checking. In the first stage, we use \texttt{GPT-4o-mini}~\citep{openai2024gpt4o} via the OpenAI API to extract up to five verifiable factual claims from each website's visible text. The extraction prompt instructs the model to identify specific, self-contained assertions of fact (e.g., statistics, dates, named entities) while excluding opinions, predictions, and vague statements. Input text is truncated to 6{,}000 characters and the model is queried at temperature $0.1$ to maximize determinism.

In the second stage, extracted claims are verified by human annotators recruited on Prolific. Annotators use a custom web-based annotation interface (see Appendix~\ref{sec:appendix_annotation}) to assess each claim against one of four verdict categories, following a scheme inspired by FEVER~\citep{thorne2018fever} and AVeriTeC~\citep{schlichtkrull2024automated}: \emph{Supported} (claim is corroborated by reliable evidence), \emph{Refuted} (claim is contradicted by reliable evidence), \emph{Not Enough Evidence} (insufficient evidence to verify), and \emph{Conflicting Evidence} (sources both support and refute the claim). Annotators are instructed to search for evidence using established sources (e.g., websites of major news, governmental, and academic organizations) and to record the URLs of all consulted sources. Each annotator evaluates claims from five articles; a $20\%$ overlap in claim assignments is maintained to compute the inter-annotator agreement.

The fact-checking study recruited $N=50$ approved annotators (after excluding 11 timed-out submissions), with a mean age of $39.8$ years ($\text{SD}=12.39$, range: 21--70), $62.0\%$ male and $38.0\%$ female. Annotators were located across multiple countries ($36.0\%$ United Kingdom, $18.0\%$ United States, $12.0\%$ Portugal, and others). The mean annotation time was $41.3$ minutes ($\text{SD}=16.1$). We compute inter-annotator agreement using Krippendorff's alpha~\citep{krippendorff2004reliability}, which accommodates missing data from incomplete overlap assignments.

We define the factual error rate as the proportion of claims labeled as \emph{Refuted} within each monthly sample. A positive correlation between this rate and the aggregate AI likelihood score would indicate truth decay.

\paragraph{Hypothesis 3: Positivity Shift.}
We measure document-level sentiment using the \texttt{cardiffnlp/twitter-roberta-base-sentiment-\allowbreak latest} model~\citep{loureiro2022timelms}, a \texttt{RoBERTa-base} model fine-tuned on 124M tweets for sentiment classification\footnote{\noindent \url{https://huggingface.co/cardiffnlp/twitter-roberta-base-sentiment-latest}}. Each document's visible text is truncated to 500 characters and tokenized with a maximum sequence length of 512 tokens. The model outputs softmax probabilities over three classes (negative, neutral, positive), and each document is assigned the label with the highest probability. We compute the rate of documents classified as positive within each monthly sample. A positive correlation between this rate and the aggregate AI likelihood score would indicate a positivity shift.

\paragraph{Hypothesis 4: Epistemic Islands.}
We measure the density of outbound hyperlinks as a proxy for epistemic connectivity. For each archived HTML page, we parse the DOM using BeautifulSoup with the \texttt{lxml} parser. We first remove structural navigation elements (\texttt{<nav>}, \texttt{<footer>}, \texttt{<header>}, \texttt{<aside>}) to isolate the article body, then count all remaining anchor tags (\texttt{<a>}) with \texttt{href} attributes. Link density is computed as the number of outbound links per 1{,}000 words of visible text; documents with fewer than 50 words are excluded. We compute the mean link density for each monthly sample. An inverse correlation between link density and the aggregate AI likelihood score would indicate epistemic island formation.

\paragraph{Hypothesis 5: Entropy Dilution.}
We measure the information density of documents using the Gzip compression ratio. For each document, we encode the extracted visible text as UTF-8, compute the size of the raw byte string, compress it using the \texttt{gzip} module from the Python standard library, and compute the ratio of compressed size to raw size. Lower compression ratios indicate higher information density (i.e., less redundancy), while higher ratios indicate more compressible, repetitive content. Documents with fewer than 100 characters of extracted text are excluded. A positive correlation between the mean compression ratio and the aggregate AI likelihood score would indicate entropy dilution.

\paragraph{Hypothesis 6: Stylistic Monoculture.}
We operationalize stylistic diversity using character-level $n$-gram similarity. For each document, we extract all character 3-grams from the first 2{,}000 characters of the lowercased visible text and represent each document as a set of unique 3-grams. Pairwise stylistic similarity between documents is computed using the Jaccard index (i.e., the ratio of the intersection to the union of the two 3-gram sets). For samples with $n \leq 100$ documents, we compute all pairwise similarities; for larger samples, we randomly draw up to 10{,}000 pairs. The mean Jaccard similarity for each monthly sample serves as the signal. A positive correlation between this signal and the aggregate AI likelihood score would indicate stylistic monoculture.

\paragraph{Statistical Testing.}
For all six hypotheses, the independent variable is the aggregate AI likelihood score (combining Pangram v3's fully AI-generated and AI-assisted fractions) aggregated at the monthly level, and the dependent variable is the corresponding monthly signal. We compute the Pearson correlation coefficient $\rho$ and its associated $p$-value. Significance is assessed at $\alpha = 0.05$.

%% file: sec/5_appendix.tex
\appendix

\section{Robustness Analysis of AI-Generated Text Detectors}
\label{sec:appendix_robustness}

We conducted a systematic robustness analysis comparing four detectors of AI-generated text---Binoculars~\citep{hans2024spotting}, Desklib~\citep{desklib2025aitextdetector}, DivEye~\citep{basani2025diversity}, and the Pangram v3 commercial API~\citep{pangram_api_docs}---across five experimental dimensions to inform our choice of detector for the main analyses.

\begin{figure}[h]
    \centering
    % Top row
    \begin{minipage}[t]{0.48\linewidth}
        \centering
        \includegraphics[width=\linewidth]{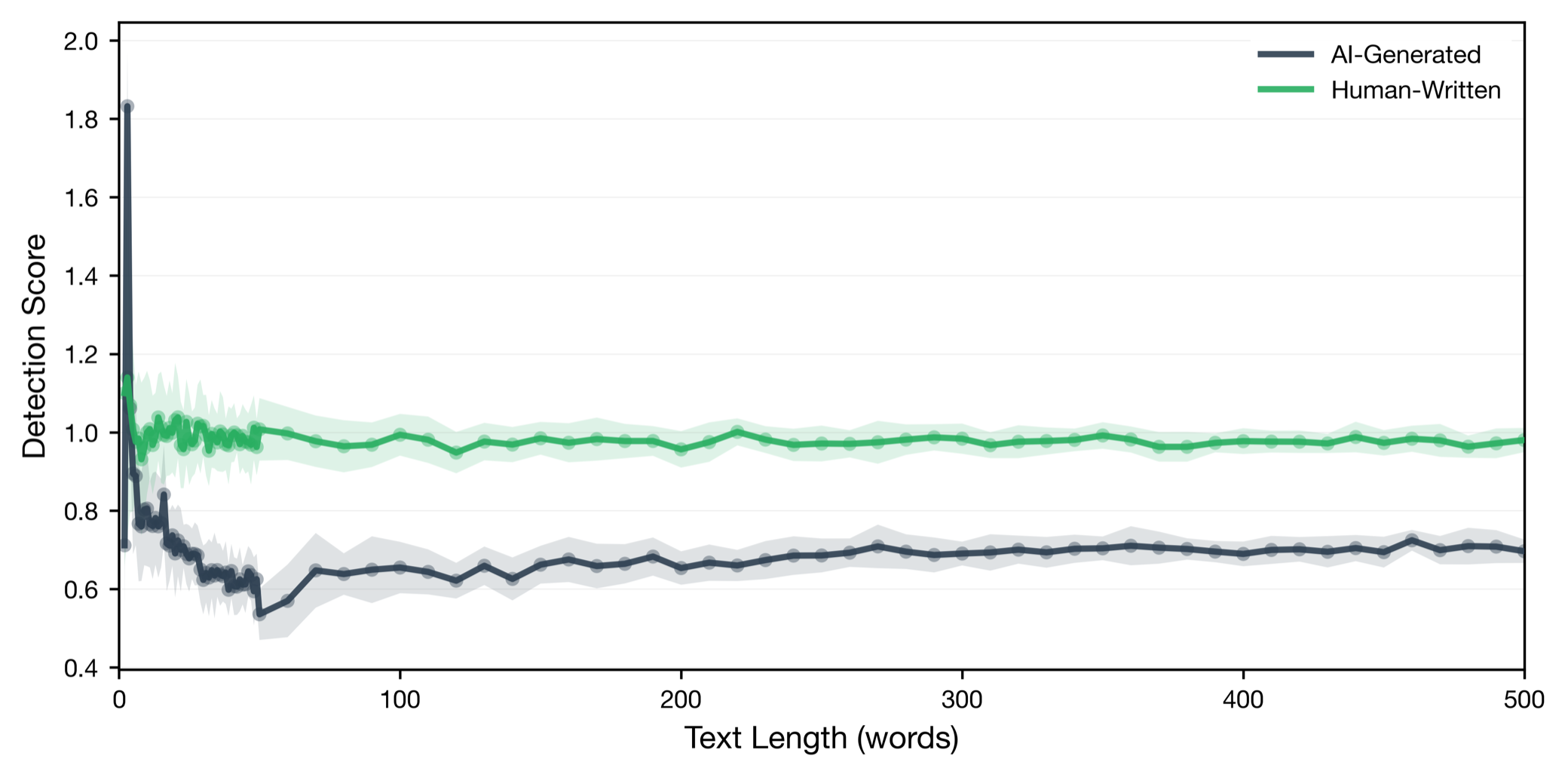}
    \end{minipage}
    \hfill
    \begin{minipage}[t]{0.48\linewidth}
        \centering
        \includegraphics[width=\linewidth]{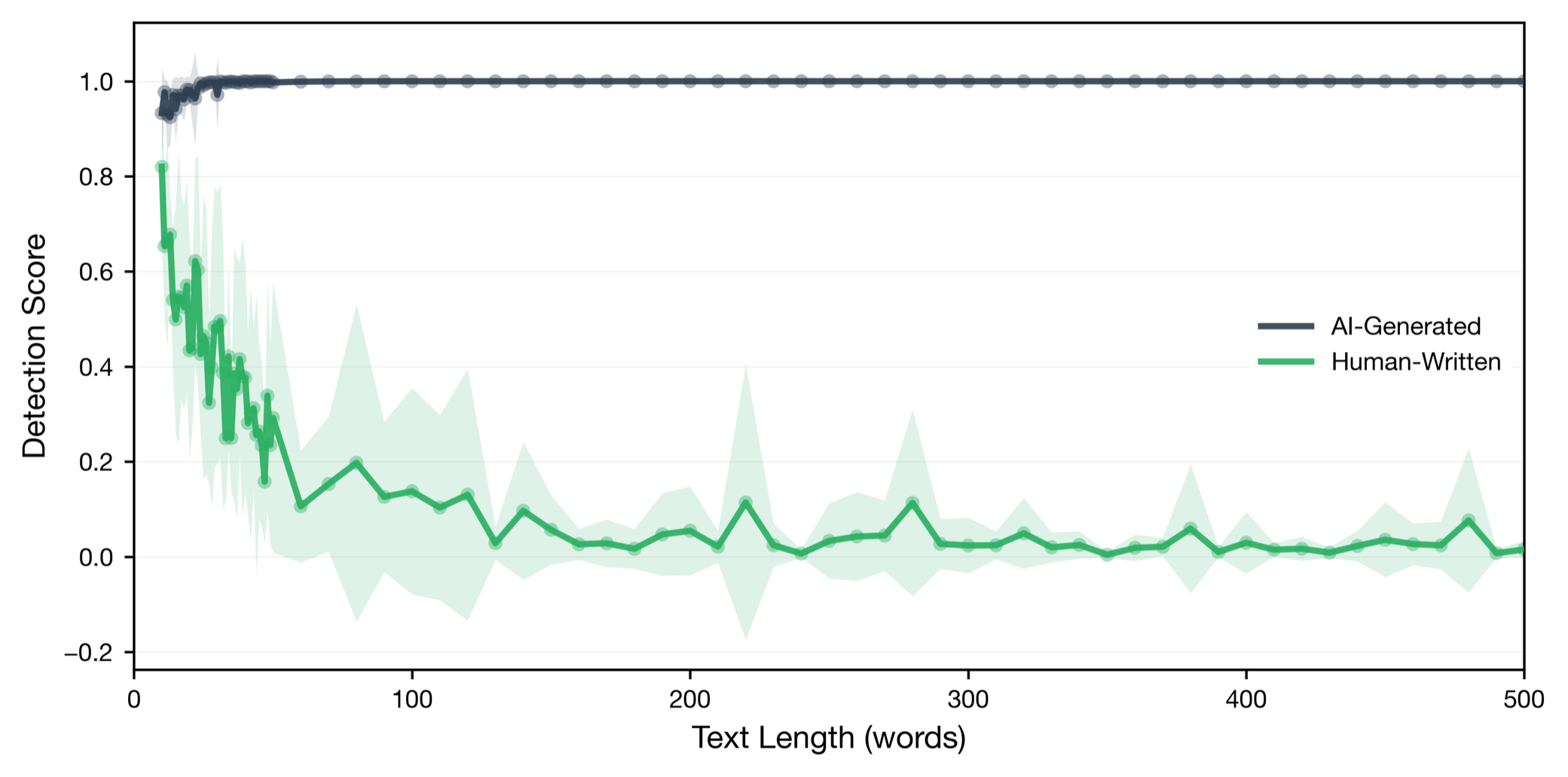}
    \end{minipage}

    \vspace{0.5em}
    
    \begin{minipage}[t]{0.48\linewidth}
        \centering
        \small 
        (a) Binoculars
    \end{minipage}
    \hfill
    \begin{minipage}[t]{0.48\linewidth}
        \centering
        \small 
        (b) Desklib
    \end{minipage}

    \vspace{0.5em}

    \begin{minipage}[t]{0.48\linewidth}
        \centering
        \includegraphics[width=\linewidth]{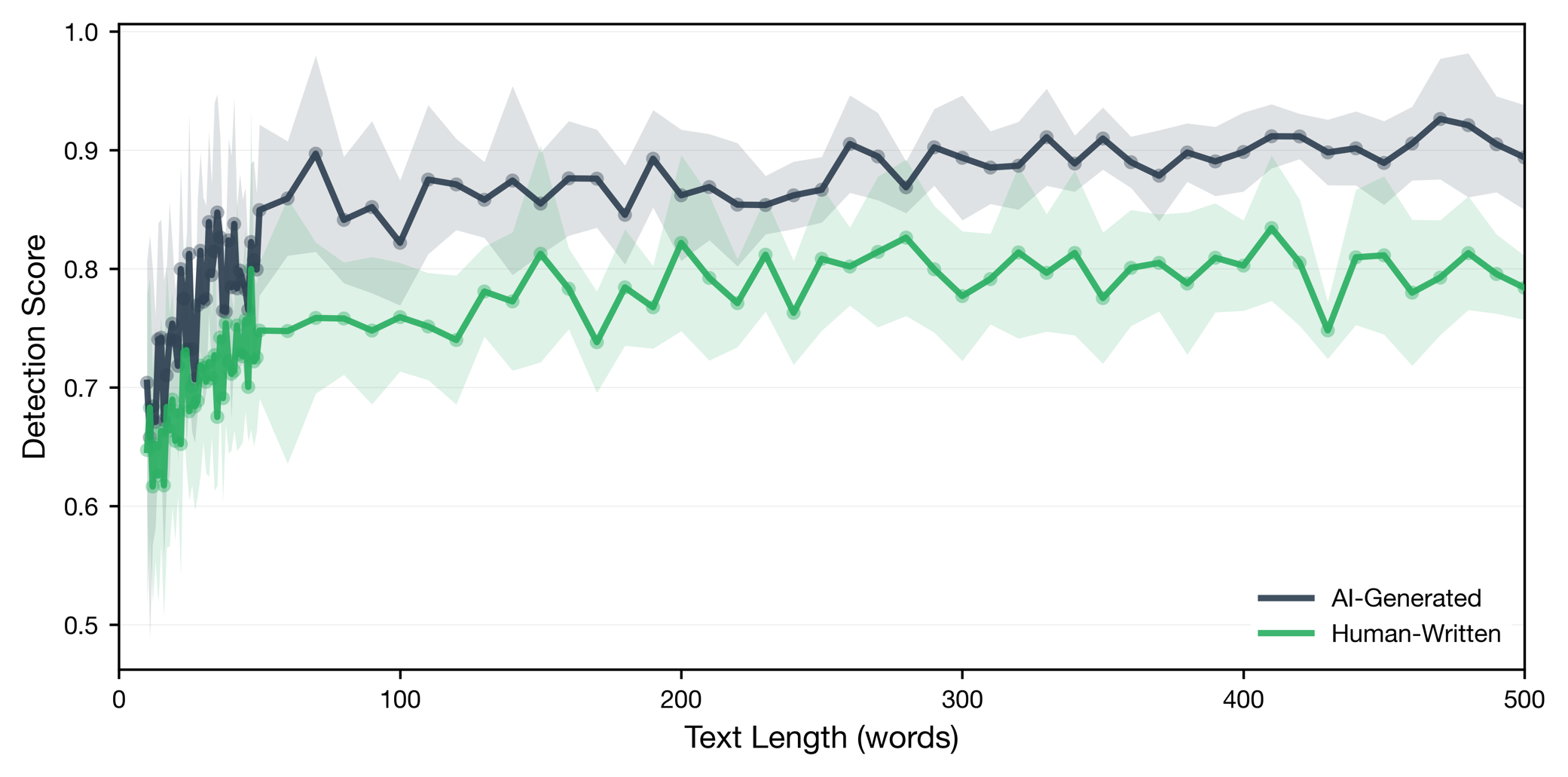}
    \end{minipage}
    \hfill
    \begin{minipage}[t]{0.48\linewidth}
        \centering
        \includegraphics[width=\linewidth]{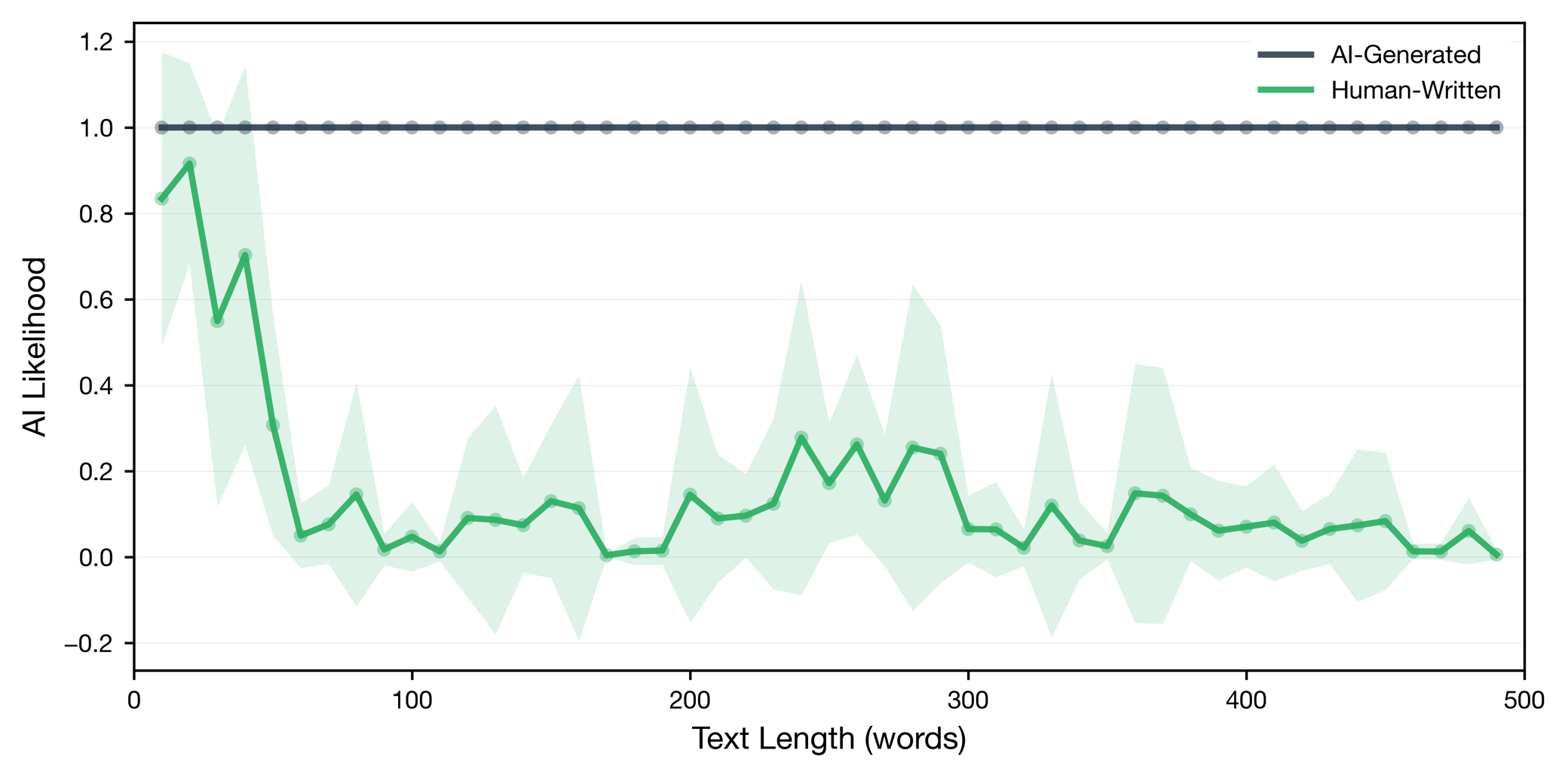}
    \end{minipage}

    \vspace{0.5em}
    
    \begin{minipage}[t]{0.48\linewidth}
        \centering
        \small 
        (c) DivEye
    \end{minipage}
    \hfill
    \begin{minipage}[t]{0.48\linewidth}
        \centering
        \small 
        (d) Pangram v3
    \end{minipage}

    \vspace{0.5em}

    \caption{\textbf{Length Sensitivity Analysis.} Detection scores for (a) Binoculars, (b) Desklib, (c) DivEye, and (d) Pangram v3 across varying text lengths. Except for DivEye, these detectors generally achieve strong performance for texts exceeding 100 words.}
    \label{fig:robustness_length}
\end{figure}

\paragraph{Text Length.}
We evaluated detection robustness across texts of various lengths, ranging from $1$ to $500$ words, using $20$ AI-generated and $20$ human-written samples per length bin. Length sensitivity curves for all detectors are shown in Figure~\ref{fig:robustness_length}, indicating the separability between AI-generated and human-written texts.

\noindent For texts of 1--10 words, Binoculars achieved a true positive rate (TPR) of 61.6\% with a false positive rate (FPR) of 35\%, rendering it unreliable for very short texts. DivEye also achieved limited separability with shorter texts. Meanwhile, Desklib and Pangram v3 scores between the distributions were smaller but still maintained separability. Performance improved substantially for texts of 11--50 words (for Binoculars: TPR: 96.6\% and FPR: 17.9\%). Binoculars, Desklib, and Pangram v3 reached near-perfect levels for texts exceeding 50 words. DivEye improved in these ranges, but registered less separability.

\FloatBarrier
\newpage

\begin{figure}[h]
    \centering
    \begin{minipage}[t]{0.48\linewidth}
        \centering
        \includegraphics[width=\linewidth]{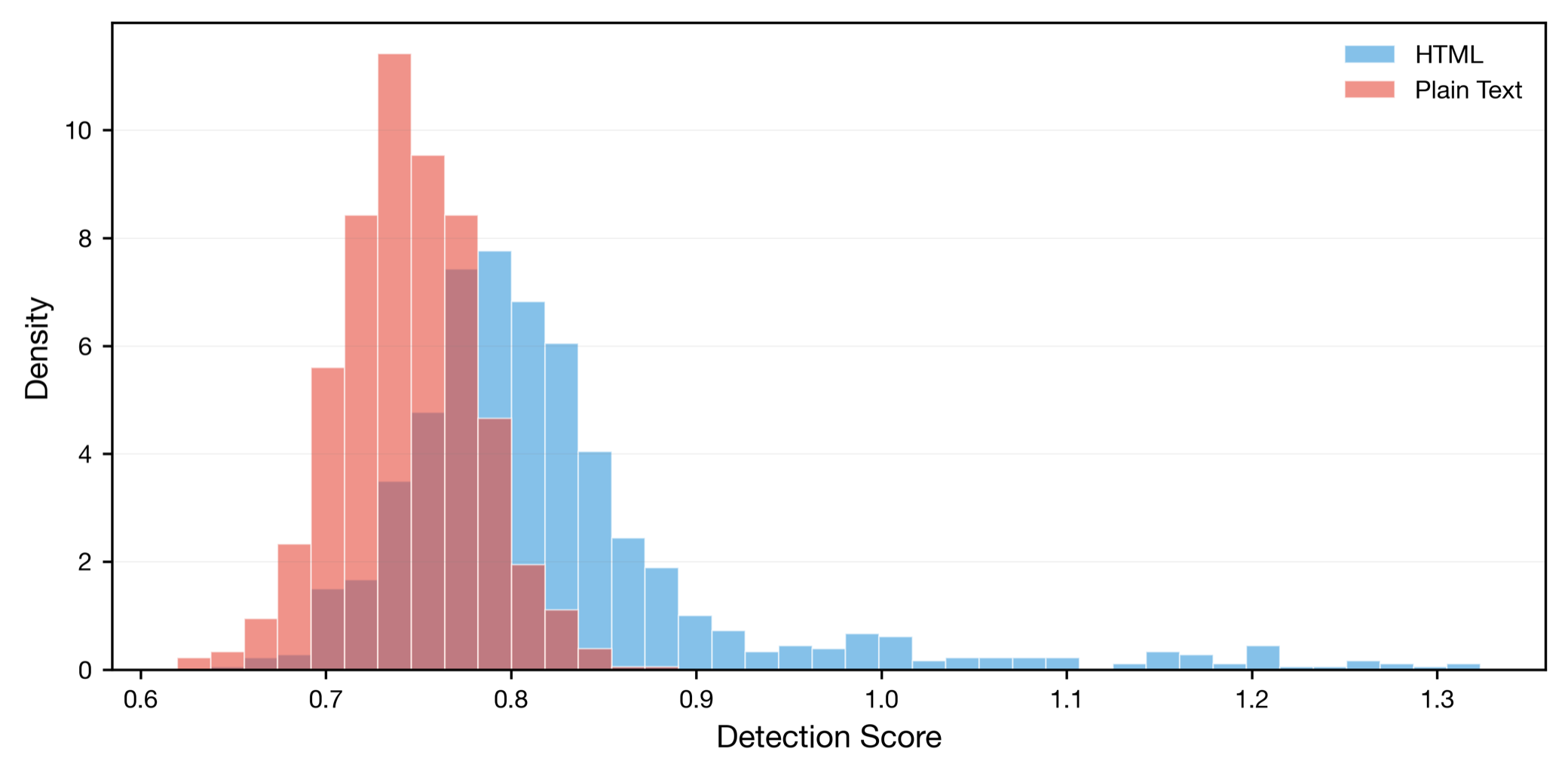}
    \end{minipage}
    \hfill
    \begin{minipage}[t]{0.48\linewidth}
        \centering
        \includegraphics[width=\linewidth]{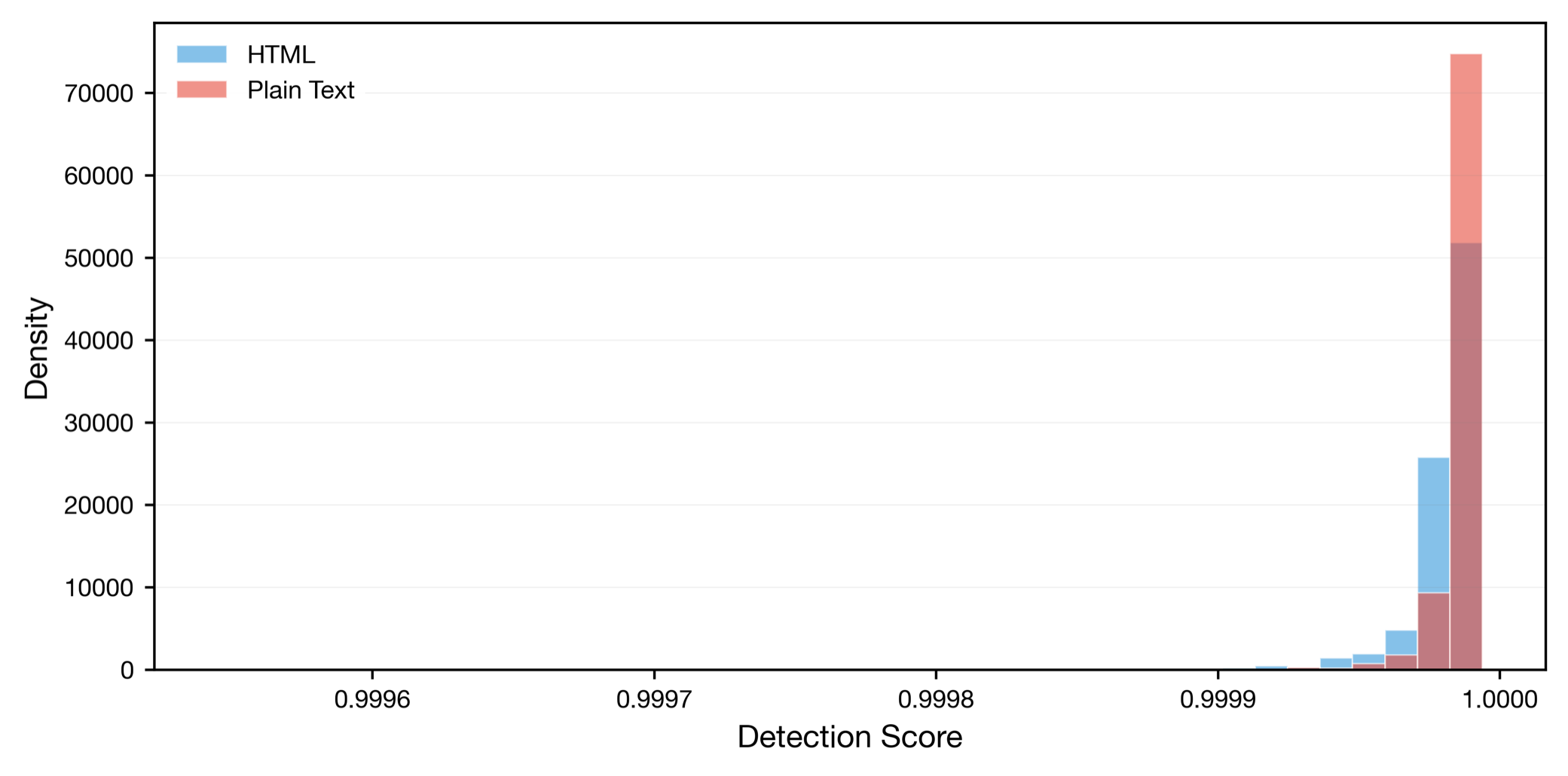}
    \end{minipage}

    \vspace{0.5em}
    
    \begin{minipage}[t]{0.48\linewidth}
        \centering
        \small 
        (a) Binoculars
    \end{minipage}
    \hfill
    \begin{minipage}[t]{0.48\linewidth}
        \centering
        \small 
        (b) Desklib
    \end{minipage}

    \vspace{0.5em}

    \centering
    \begin{minipage}[t]{0.48\linewidth}
        \centering
        \includegraphics[width=\linewidth]{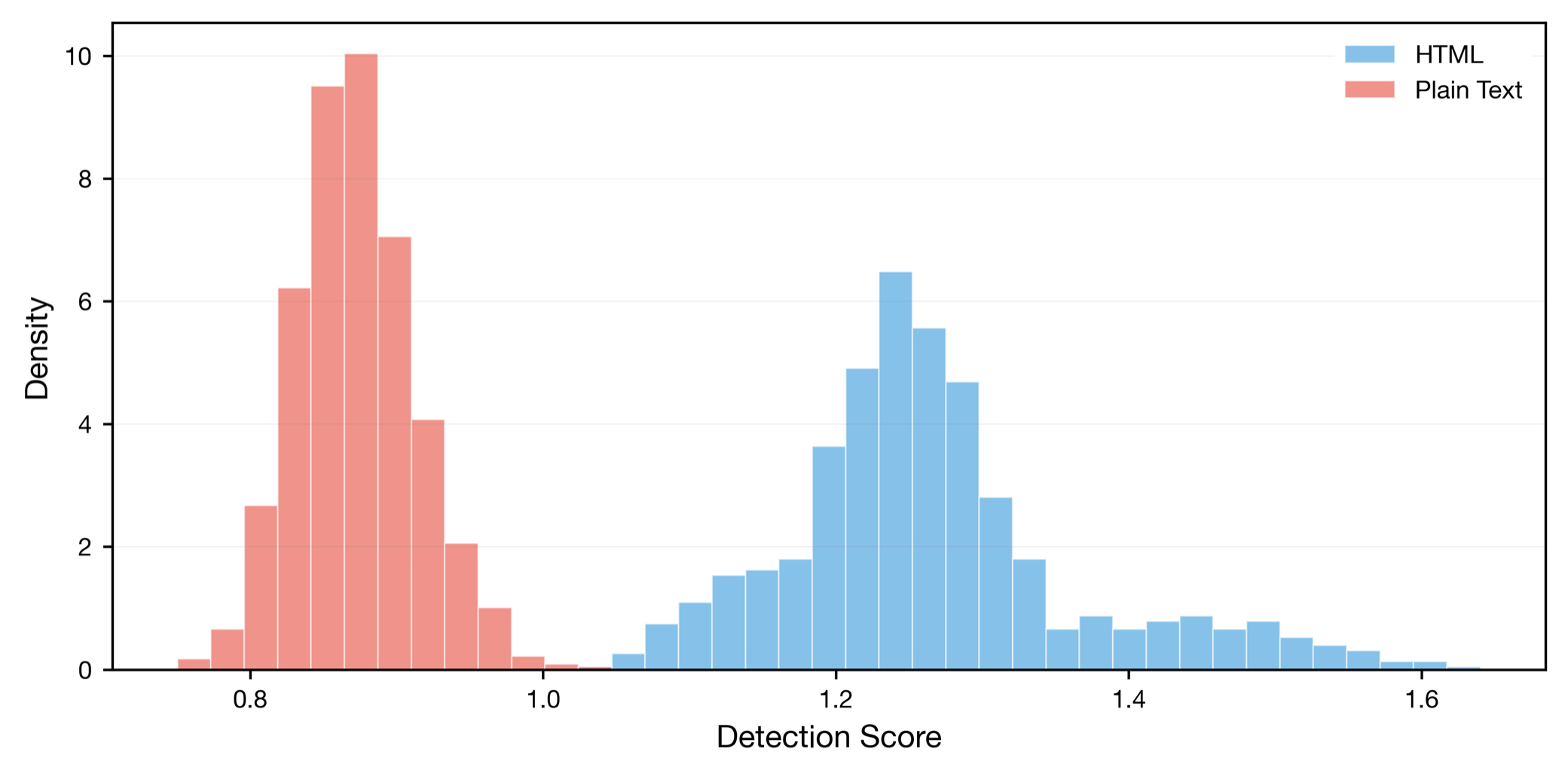}
    \end{minipage}
    \hfill
    \begin{minipage}[t]{0.48\linewidth}
        \centering
        \includegraphics[width=\linewidth]{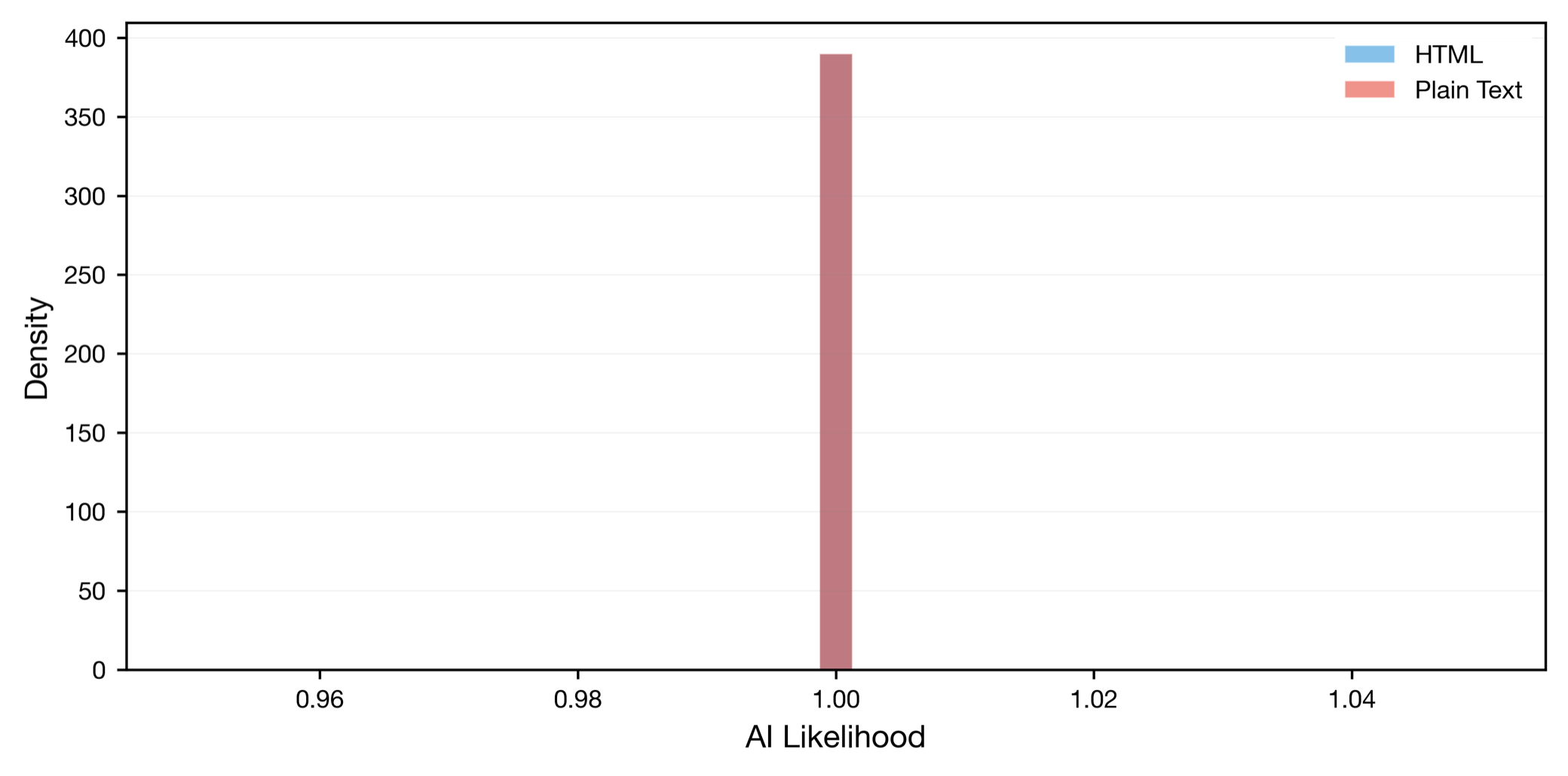}
    \end{minipage}

    \vspace{0.5em}
    
    \begin{minipage}[t]{0.48\linewidth}
        \centering
        \small 
        (c) DivEye
    \end{minipage}
    \hfill
    \begin{minipage}[t]{0.48\linewidth}
        \centering
        \small 
        (d) Pangram v3
    \end{minipage}

    \vspace{0.5em}
    
    \caption{\textbf{HTML vs. Plain Text Robustness.} Detection score distributions for AI-generated text presented in plain text versus HTML-embedded format, for (a) Binoculars, (b) Desklib, (c) DivEye, and (d) Pangram v3. Binoculars and DivEye exhibit a substantial drop in detection rate when text is embedded in HTML, Desklib exhibits some drop, and Pangram v3 exhibits minimal drop in scores.}
    \label{fig:robustness_html}
\end{figure}

\paragraph{HTML vs. Plain Text.}
We tested whether embedding AI-generated text within an HTML document structure affects detection accuracy. Identical AI-generated texts (produced by GPT-4o) were presented to all four detectors in both plain text and HTML-embedded formats. In Figure~\ref{fig:robustness_html}, we plot scores of these two distributions (plain text in red and HTML-embedded text in blue) for each tested model. For our use case, it would be ideal if the two distributions maintained a complete overlap, which would suggest that the distribution of AI-generated articles would generally register the same scores no matter whether wrapped in HTML or presented as plain text.

\noindent DivEye showed the largest divergence, wherein the two distributions had no overlap. Next, Binoculars showed partial divergence: the mean score shifted from $0.745$ ($100\%$ flag rate, if detection threshold were applied) in plain text to $0.823$ ($88.6\%$ flag rate, if detection threshold were applied) when HTML-embedded, an $11.4$ percentage point drop in detection rate. Next, Desklib registered a minor shift between the two distributions; still, the scores for both distributions remained fully above $0.9999$ (where $1.0$ is the maximum possible score for AI-generated texts), and so this is not a major concern. Finally, Pangram v3 registered a complete perfect overlap between the two distributions.

\FloatBarrier
\newpage

\begin{figure}[h]
    \centering
    \begin{minipage}[t]{0.48\linewidth}
        \centering
        \includegraphics[width=\linewidth]{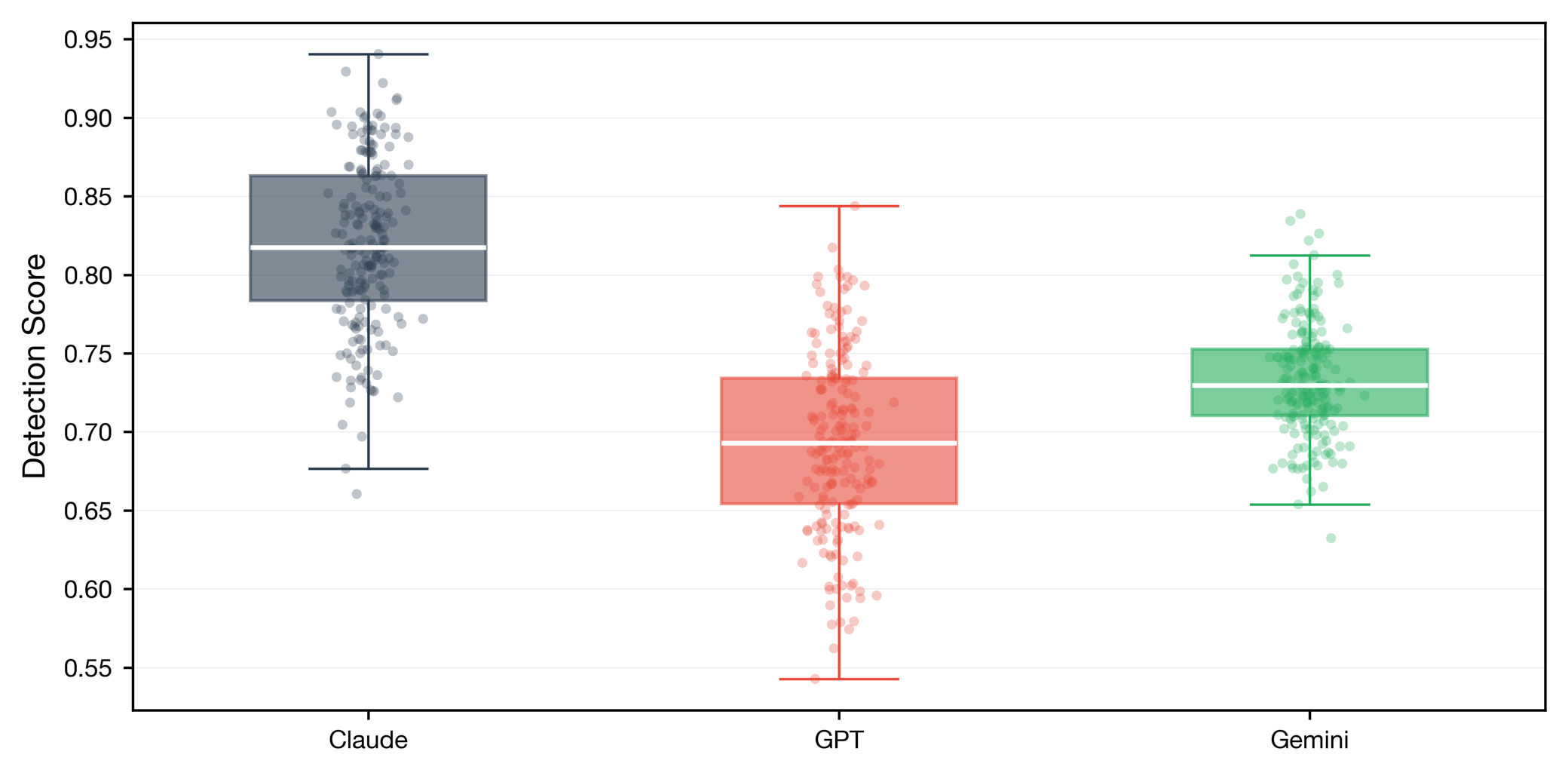}
    \end{minipage}
    \hfill
    \begin{minipage}[t]{0.48\linewidth}
        \centering
        \includegraphics[width=\linewidth]{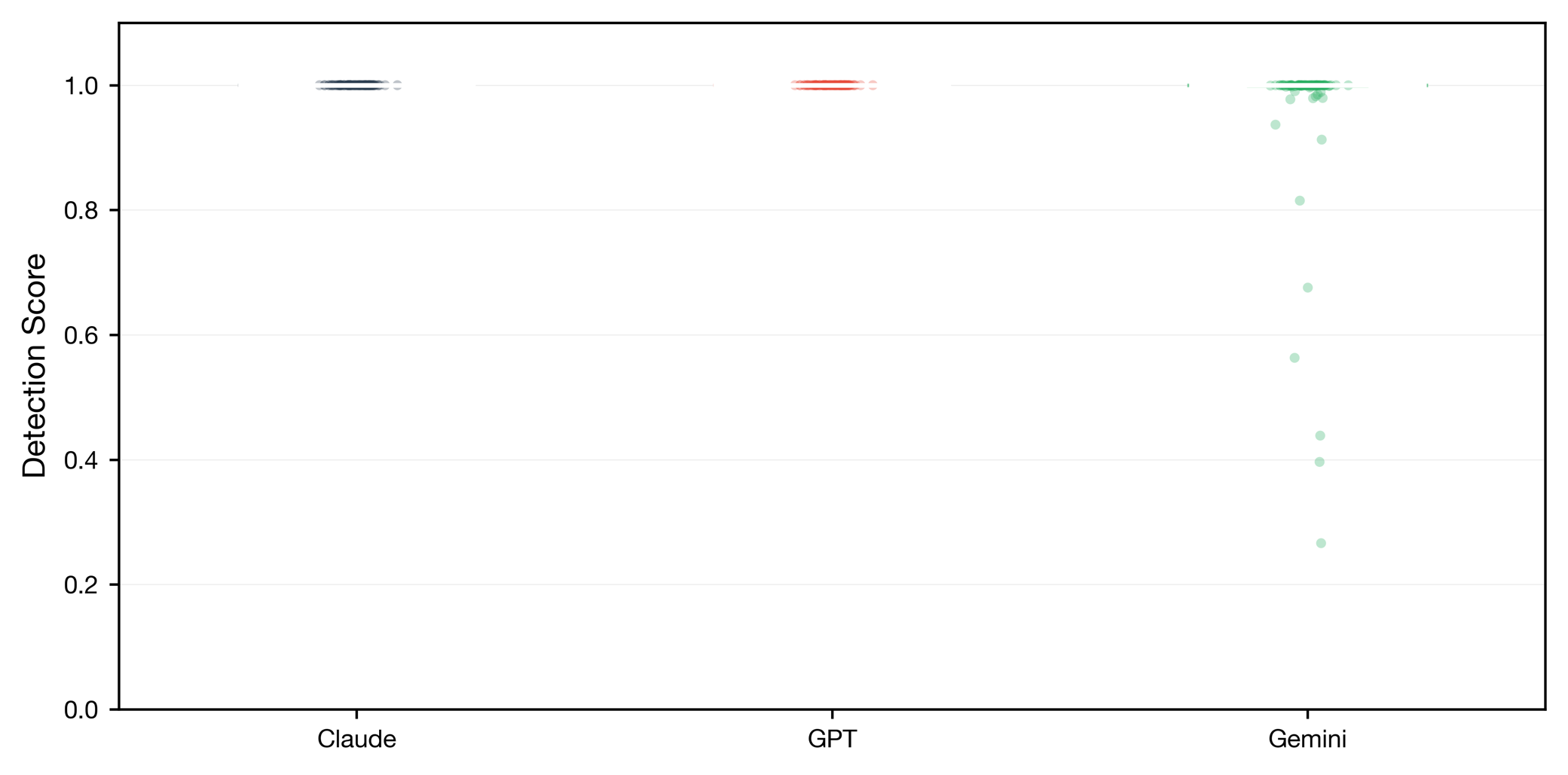}
    \end{minipage}

    \vspace{0.5em}
    
    \begin{minipage}[t]{0.48\linewidth}
        \centering
        \small 
        (a) Binoculars
    \end{minipage}
    \hfill
    \begin{minipage}[t]{0.48\linewidth}
        \centering
        \small 
        (b) Desklib
    \end{minipage}

    \vspace{0.5em}

    \centering
    \begin{minipage}[t]{0.48\linewidth}
        \centering
        \includegraphics[width=\linewidth]{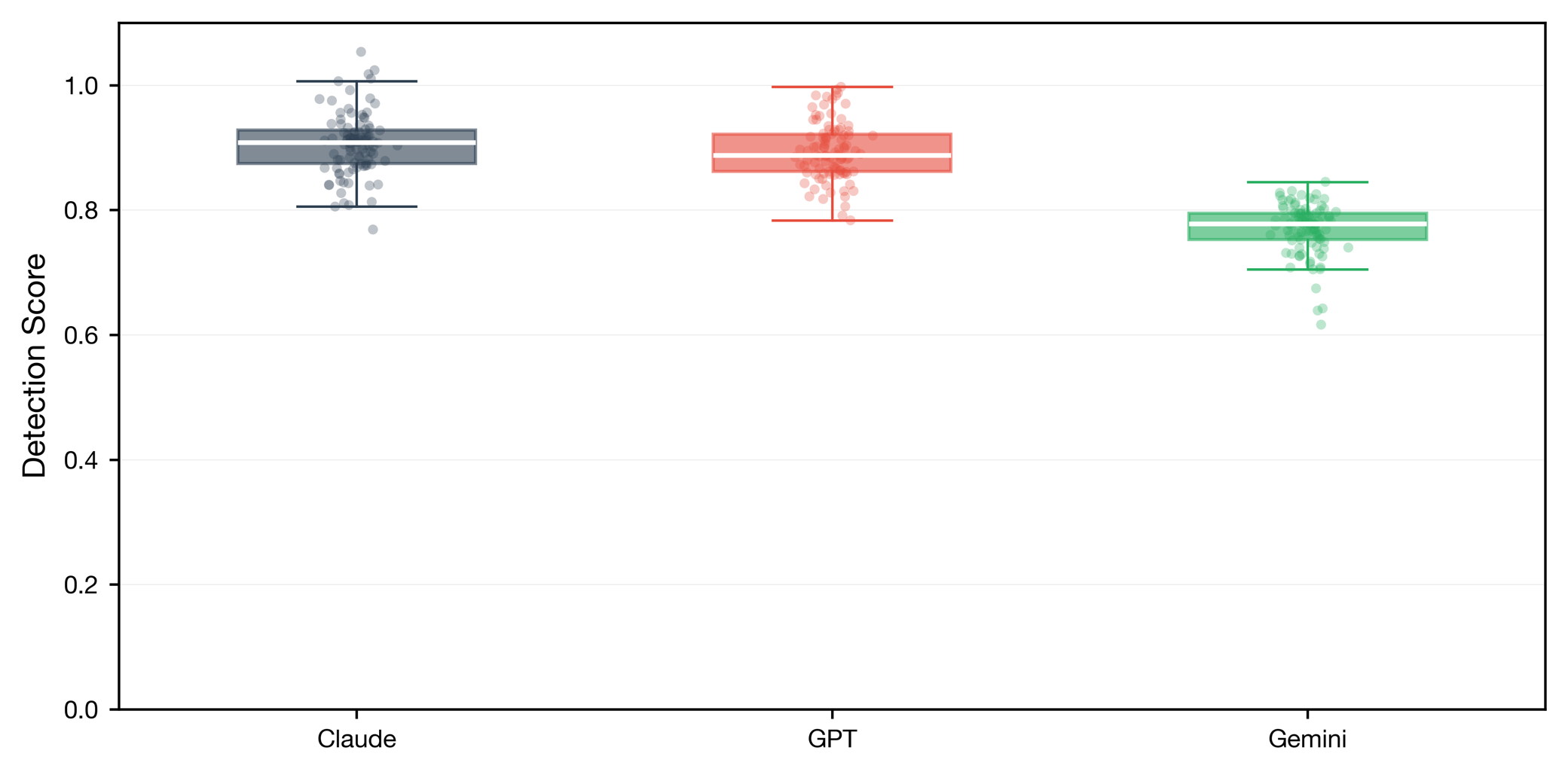}
    \end{minipage}
    \hfill
    \begin{minipage}[t]{0.48\linewidth}
        \centering
        \includegraphics[width=\linewidth]{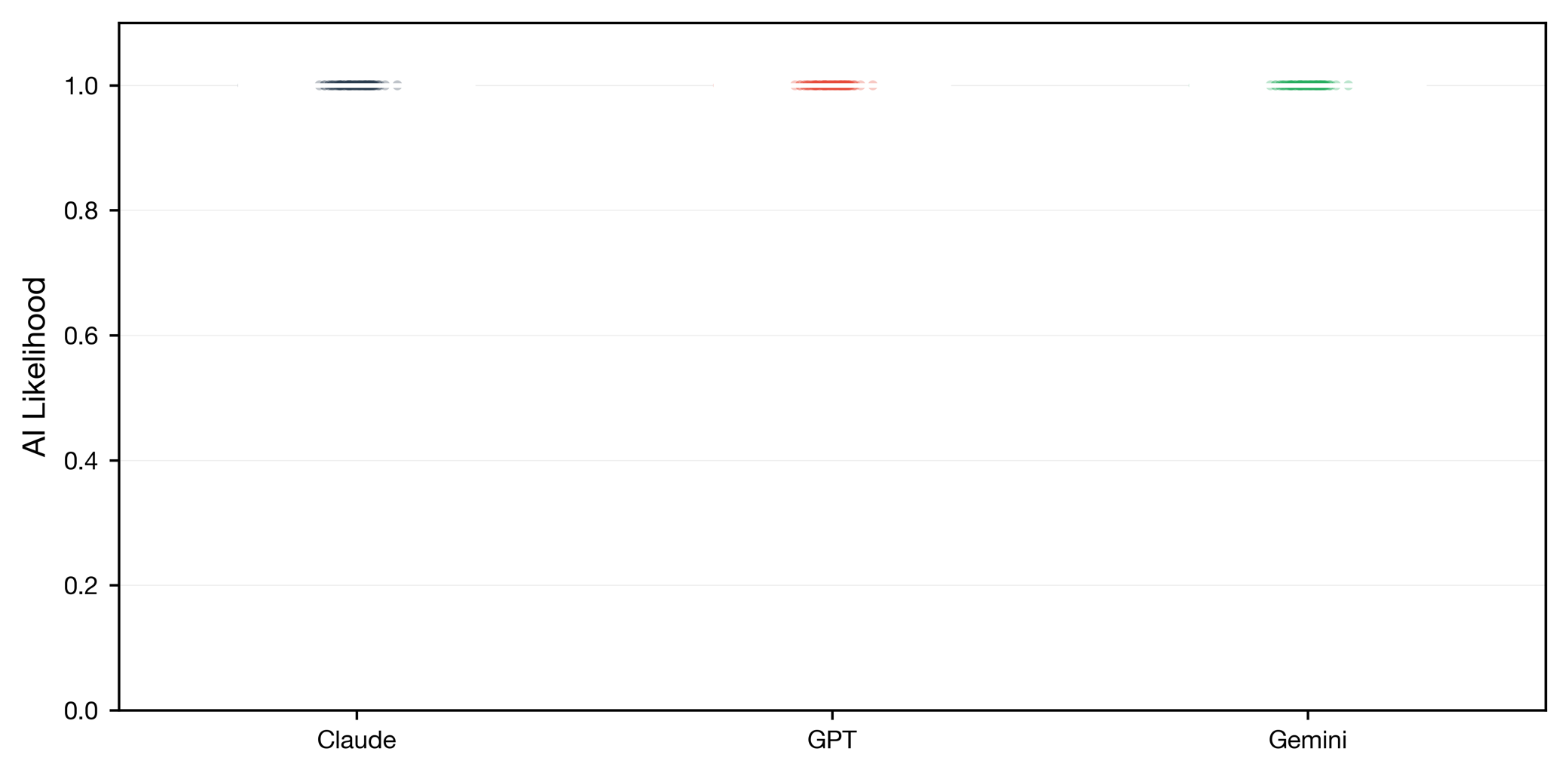}
    \end{minipage}

    \vspace{0.5em}
    
    \begin{minipage}[t]{0.48\linewidth}
        \centering
        \small 
        (c) DivEye
    \end{minipage}
    \hfill
    \begin{minipage}[t]{0.48\linewidth}
        \centering
        \small 
        (d) Pangram v3
    \end{minipage}

    \vspace{0.5em}
    
    \caption{\textbf{Model Family Robustness.} Detection score distributions for text generated by GPT-4o, Claude, and Gemini, for (a) Binoculars, (b) Desklib, (c) DivEye, and (d) Pangram v3. Binoculars performs differenly on Claude, GPT, and Gemini model families; Desklib and Pangram v3 have a similar performance across the board.}
    \label{fig:robustness_models}
\end{figure}

\paragraph{Model Family.}
We evaluated model family robustness of the tested detectors across three leading LLM families: GPT-4o (OpenAI), Claude (Anthropic), and Gemini (Google DeepMind). Shown in Figure~\ref{fig:robustness_models} are the distributions of scores for these model families registered by each of the tested detectors.

\noindent For Binoculars, GPT was easiest to detect, Gemini was second, and Claude was the hardest to detect. For DivEye, by contrast, Gemini was the hardest to detect, while Claude and GPT appeared similarly difficult. Compared to Claude and GPT, Desklib registered minor difficulty with Gemini in some cases, but generally performed similarly across the three. Pangram performed consistently across all three model families.

\FloatBarrier
\newpage

\begin{figure}[h]
    \centering
    \begin{minipage}[t]{0.48\linewidth}
        \centering
        \includegraphics[width=\linewidth]{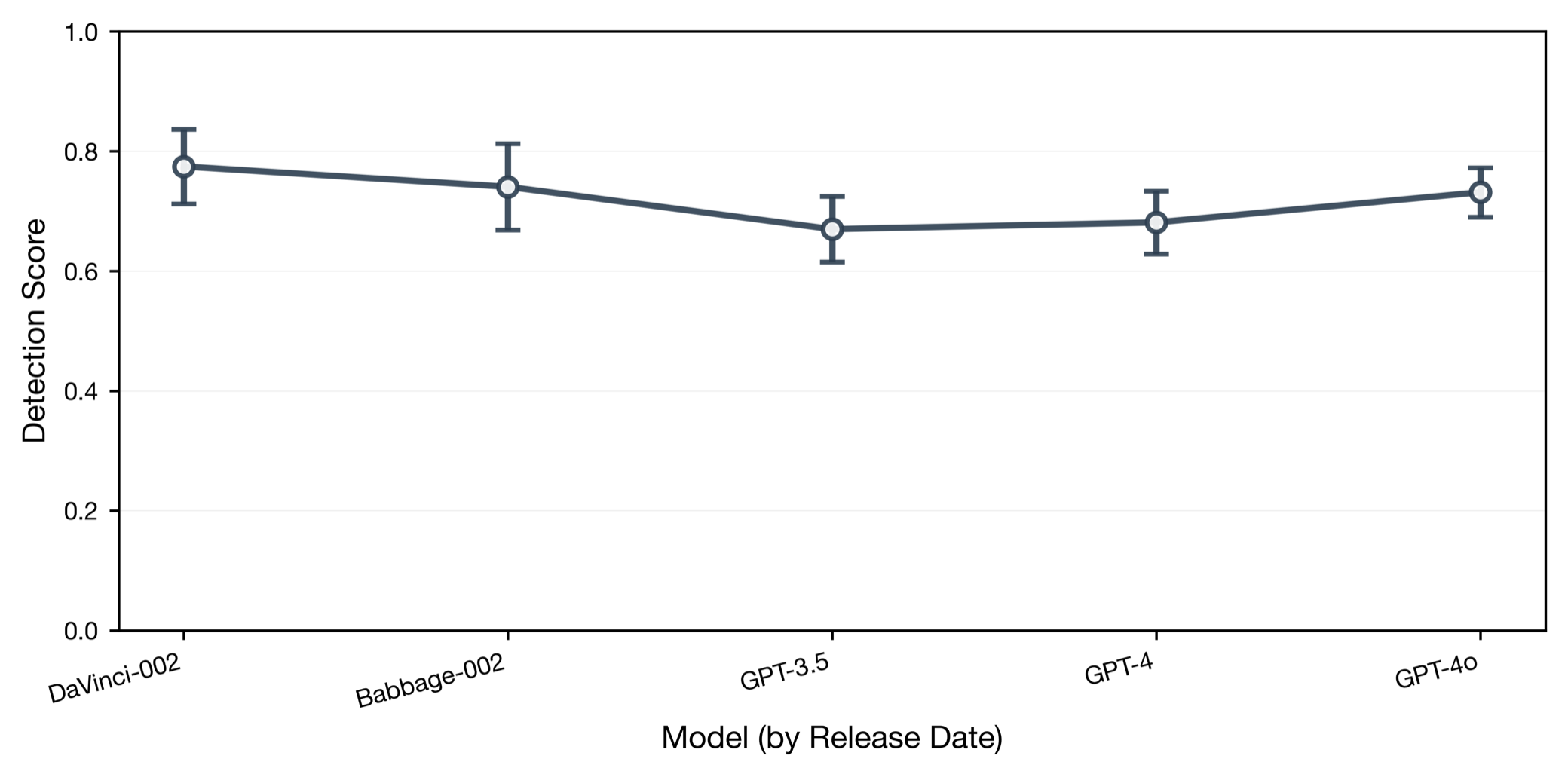}
    \end{minipage}
    \hfill
    \begin{minipage}[t]{0.48\linewidth}
        \centering
        \includegraphics[width=\linewidth]{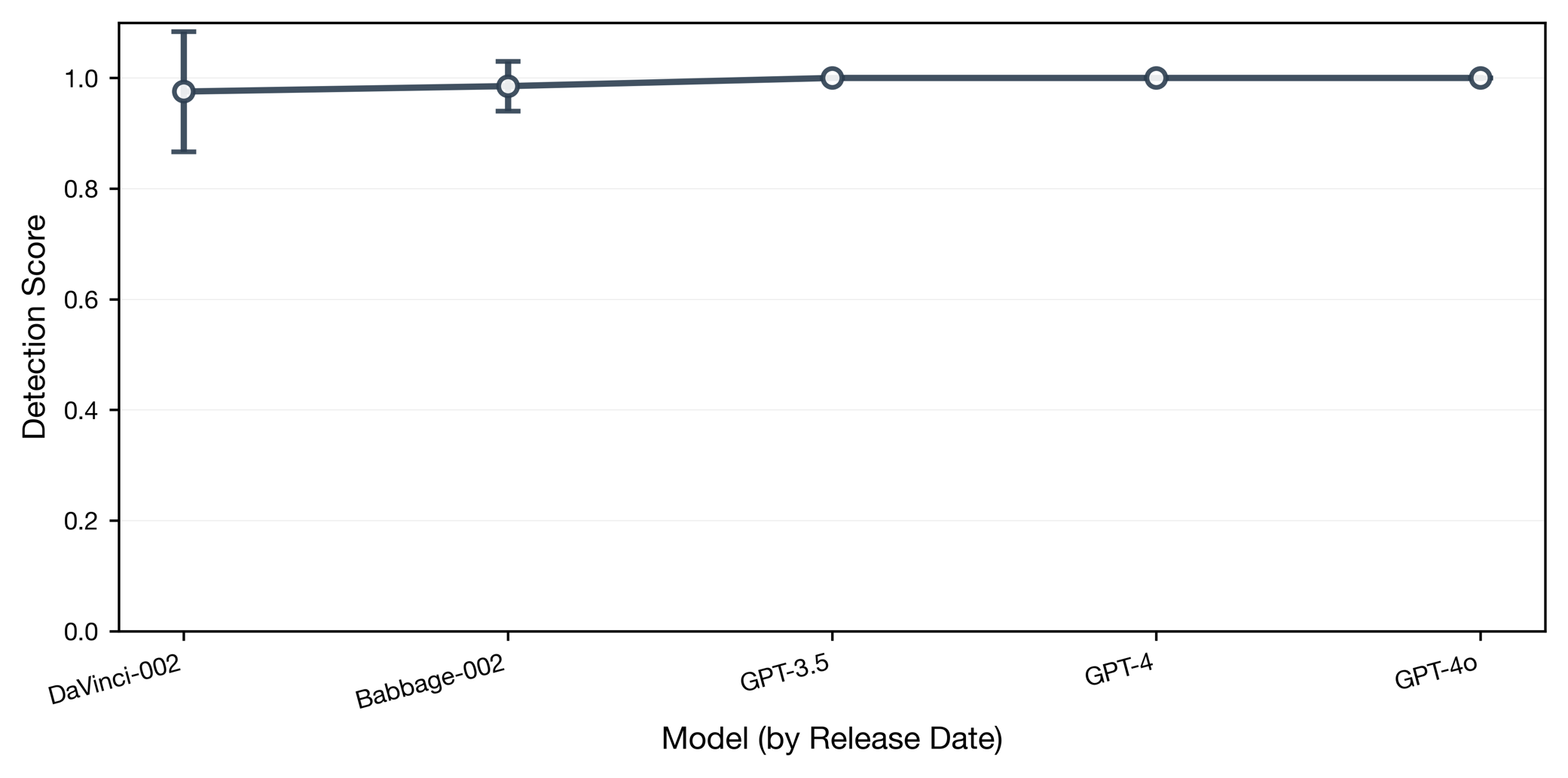}
    \end{minipage}

    \vspace{0.5em}
    
    \begin{minipage}[t]{0.48\linewidth}
        \centering
        \small 
        (a) Binoculars
    \end{minipage}
    \hfill
    \begin{minipage}[t]{0.48\linewidth}
        \centering
        \small 
        (b) Desklib
    \end{minipage}

    \vspace{0.5em}

    \centering
    \begin{minipage}[t]{0.48\linewidth}
        \centering
        \includegraphics[width=\linewidth]{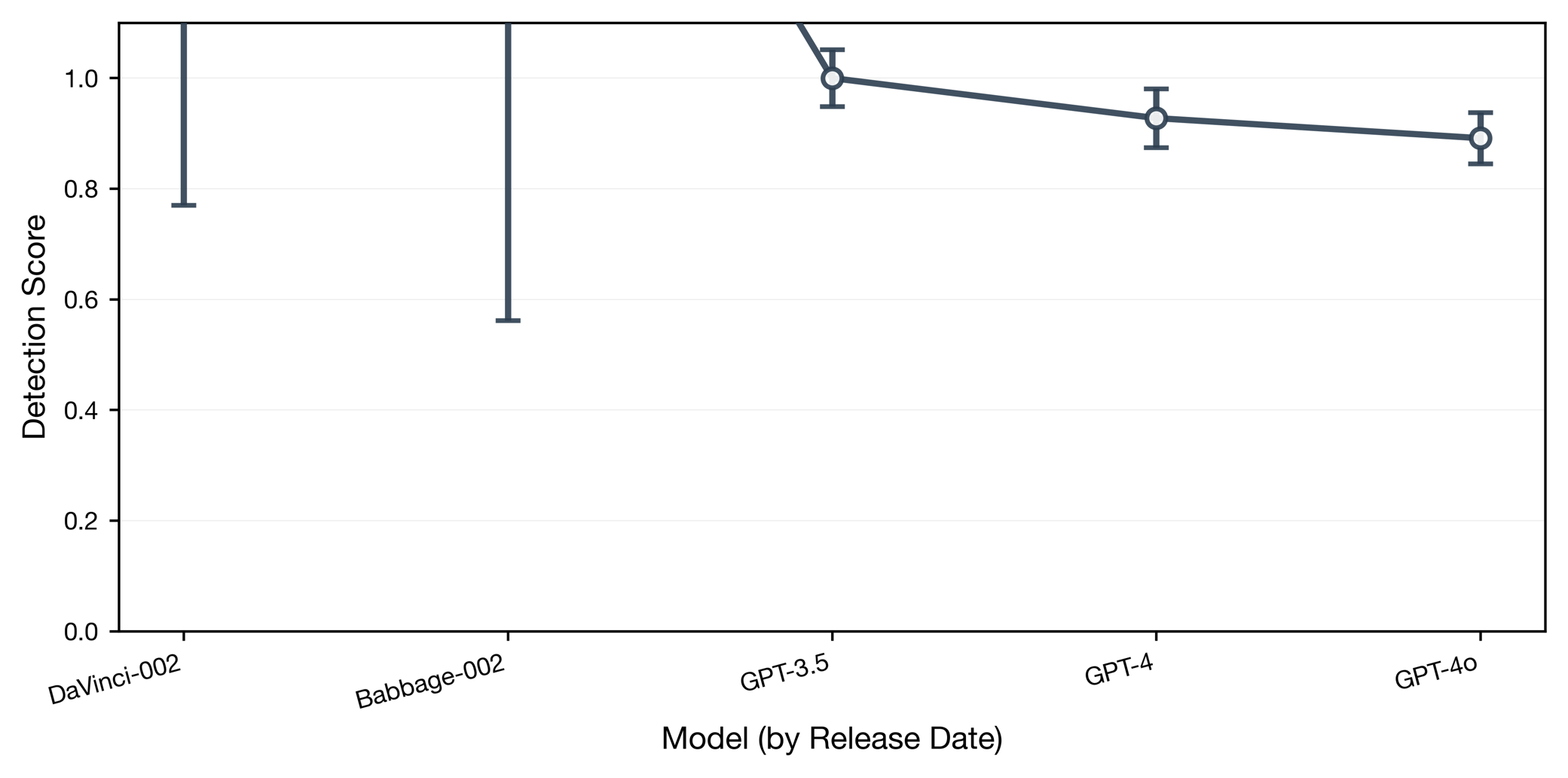}
    \end{minipage}
    \hfill
    \begin{minipage}[t]{0.48\linewidth}
        \centering
        \includegraphics[width=\linewidth]{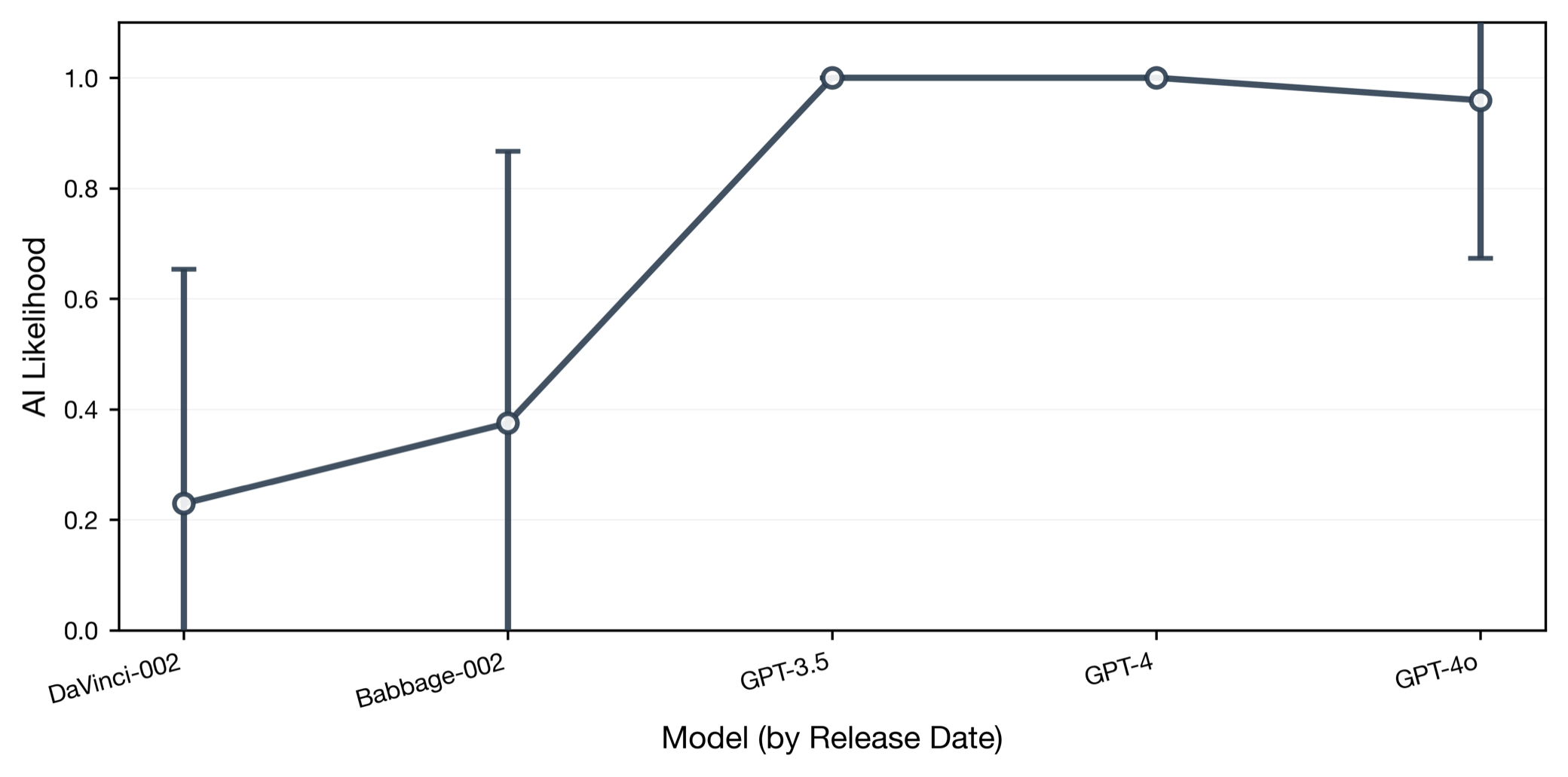}
    \end{minipage}

    \vspace{0.5em}
    
    \begin{minipage}[t]{0.48\linewidth}
        \centering
        \small 
        (c) DivEye
    \end{minipage}
    \hfill
    \begin{minipage}[t]{0.48\linewidth}
        \centering
        \small 
        (d) Pangram v3
    \end{minipage}

    \vspace{0.5em}
    
    \caption{\textbf{Model Version Robustness.} Detection scores across OpenAI model generations for (a) Binoculars, (b) Desklib, (c) DivEye, and (d) Pangram v3. Binoculars and Desklib perform consistently across the board; DivEye and Pangram v3 perform differently on completion models (\texttt{davinci-002} and \texttt{babbage-002}) compared to instruction-trained models (GPT-3.5, GPT-4, and GPT-4o).}
    \label{fig:robustness_generation}
\end{figure}

\paragraph{Model Version.}
We evaluated historical model version robustness across five generations of GPT-family models (OpenAI): \texttt{davinci-002}, \texttt{babbage-002}, \texttt{GPT-3.5}, \texttt{GPT-4}, and \texttt{GPT-4o}. The average scores, where the models are ordered from left to right by their date of release, are shown in Figure~\ref{fig:robustness_generation}.

\noindent Binoculars and DivEye provide consistent results across all model versions. Pangram v3 does not detect texts generated by the two oldest model versions, \texttt{davinci-002} and \texttt{babbage-002}, which are old completion models. DivEye exhibited a similar inconsistency for completion-only models, but also registered a slightly decreasing performance from the later instruction-tuned \texttt{GPT-3.5} through \texttt{GPT-4o}.

\noindent While temporal consistency and robustness is generally important for best results, it is critical only for the instruction-tuned models which were available to the public through chatbot interfaces. The previous completion-only models were mostly used through APIs and generally discouraged for usage shortly after the public release of ChatGPT, and hence do not constitute a major concern.

\FloatBarrier
\newpage

\paragraph{Language.}
We evaluated language robustness of the four tested detectors on AI-generated and human-written text across multiple languages: English, French, Japanese, Mandarin, and Spanish. Language-specific score distributions are shown in Figure~\ref{fig:robustness_language}. 

\noindent \textbf{Binoculars} scores yielded nearly nonoverlapping distributions for English, French, Mandarin, and Spanish; however, the ranges of these distributions (and with them, the functional threshold) shifted. Japanese was not separable.

\noindent \textbf{Desklib} scores were bimodal and separable for English and Spanish. For French, Japanese, and Mandarin, the model behaved differently, with partial overlap.

\noindent \textbf{DivEye} scores yielded nearly nonoverlapping distributions for English. For all other languages we tested (French, Japanese, Mandarin, and Spanish), the score distributions were largely not separable, and their ranges varied.

\noindent \textbf{Pangram v3} scores yielded bimodal distributions for all tested languages. Notably, the detector behaved consistently regardless of the language of the input text.

\begin{figure}[H]
    \captionsetup{justification=centering}
    \centering

    \begin{subfigure}{\linewidth}
        \centering
        \captionsetup{justification=centering}
        \includegraphics[width=0.9\linewidth]{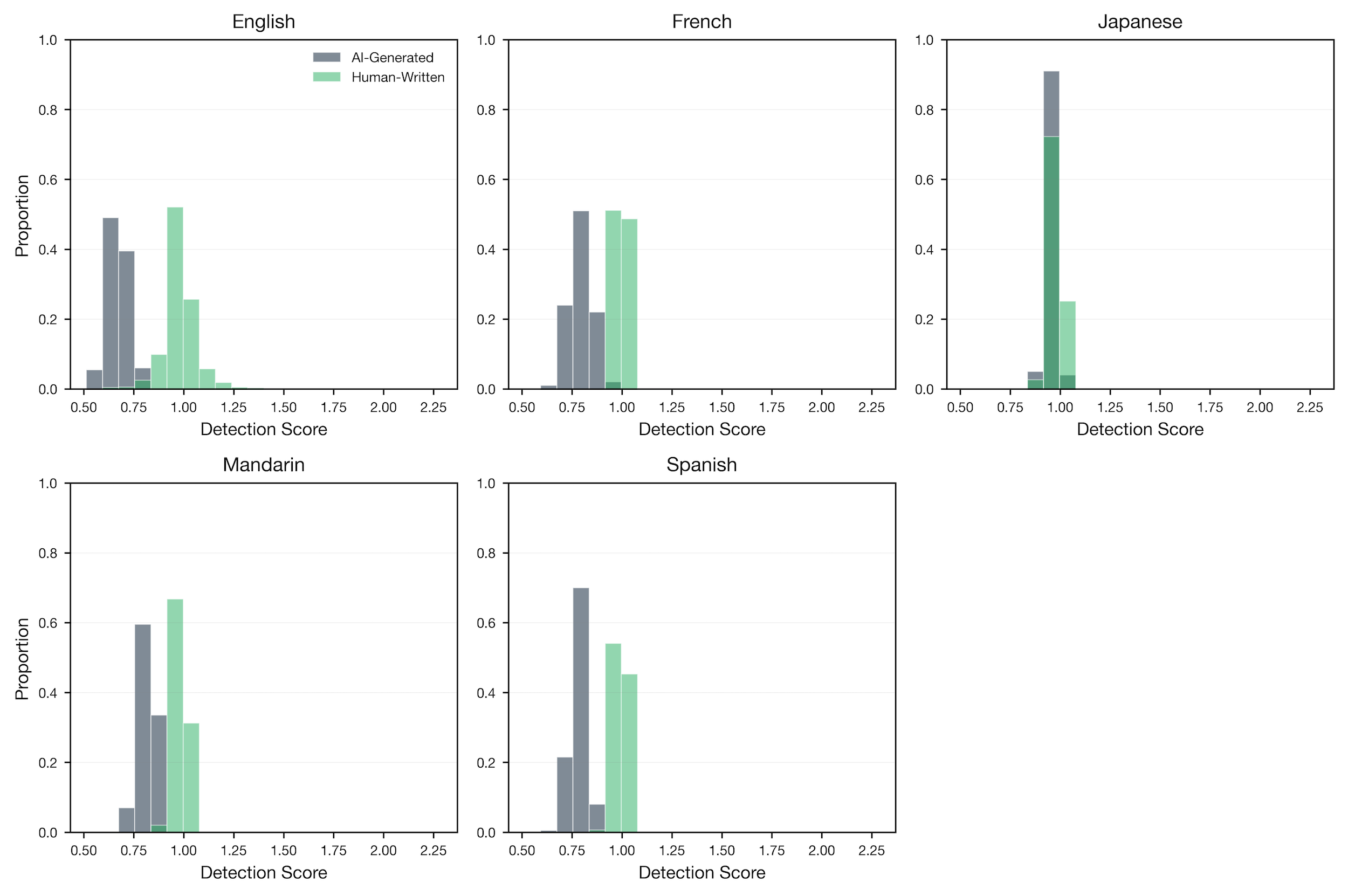}
        \caption{Binoculars}
    \end{subfigure}

    \caption{\textbf{Language Robustness.} Comparison of detector score distributions across languages for Binoculars, Desklib, DivEye, and Pangram v3 (1/3).}
    \label{fig:robustness_language}
\end{figure}

\begin{figure}[p]
    \ContinuedFloat
    \centering
    
    \begin{subfigure}{\linewidth}
        \centering
        \captionsetup{justification=centering}
        \includegraphics[width=0.9\linewidth]{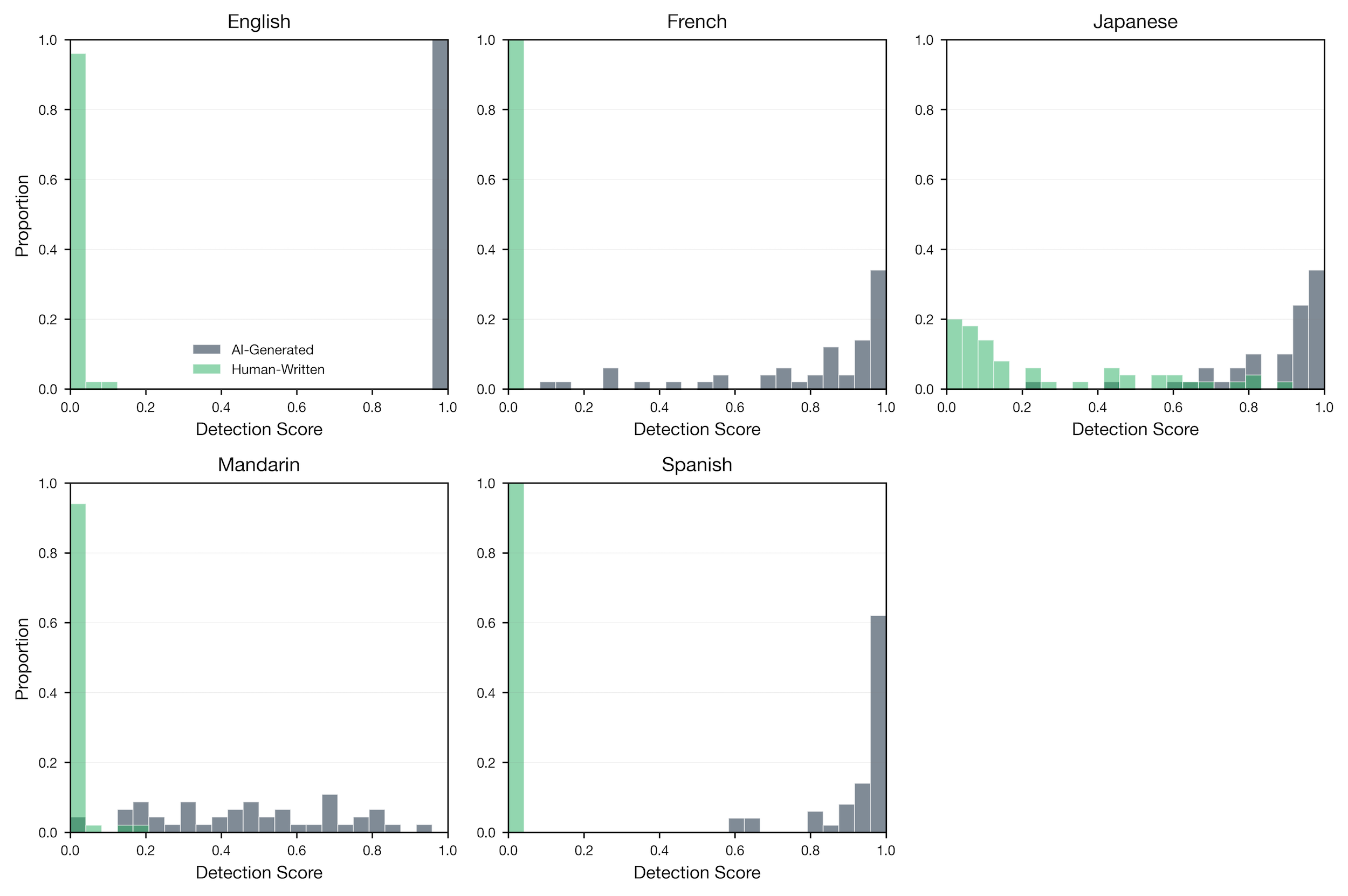}
        \caption{Desklib}
    \end{subfigure}

    \vspace{0.75em}

    \begin{subfigure}{\linewidth}
        \centering
        \captionsetup{justification=centering}
        \includegraphics[width=0.9\linewidth]{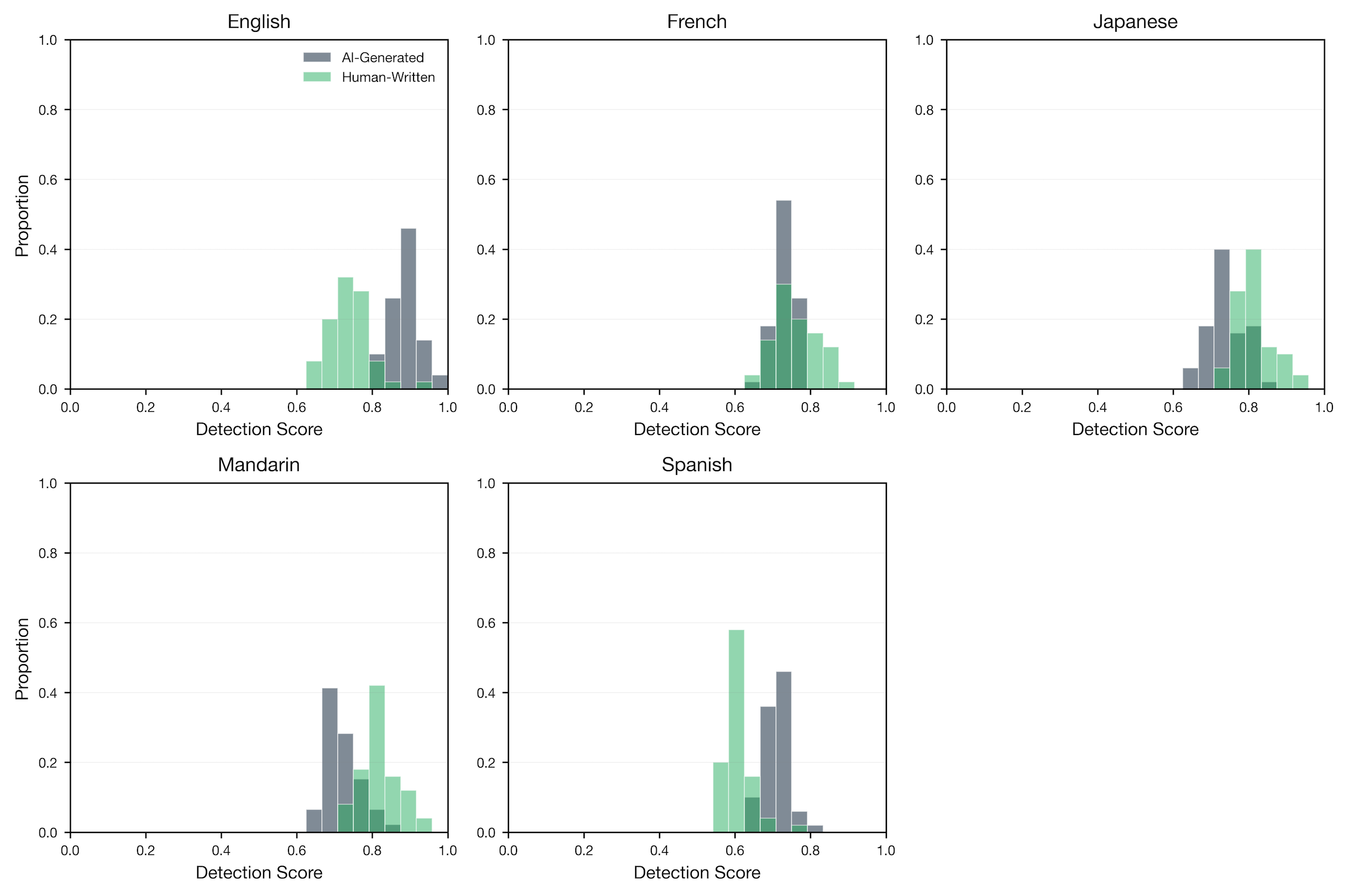}
        \caption{DivEye}
    \end{subfigure}

    \caption[]{\textbf{Language Robustness.} Comparison of detector score distributions across languages for Binoculars, Desklib, DivEye, and Pangram v3 (2/3).}
\end{figure}

\begin{figure}[t]
    \ContinuedFloat
    \centering

    \begin{subfigure}{\linewidth}
        \centering
        \captionsetup{justification=centering}
        \includegraphics[width=0.9\linewidth]{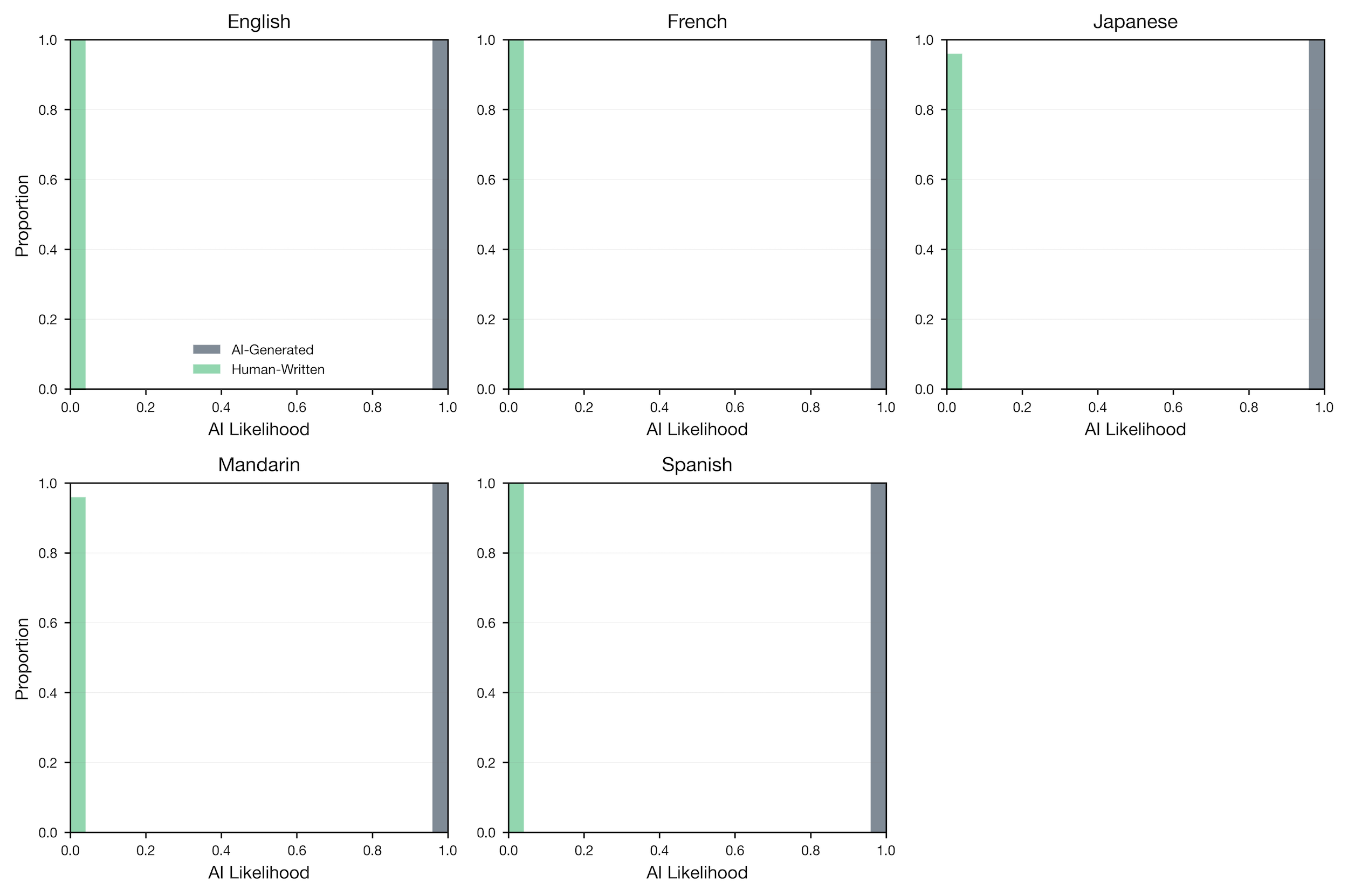}
        \caption{Pangram v3}
    \end{subfigure}

    \caption[]{\textbf{Language Robustness.} Comparison of detector score distributions across languages for Binoculars, Desklib, DivEye, and Pangram v3 (3/3).}
\end{figure}

\FloatBarrier
\newpage

\paragraph{Summary.}

Based on this evaluation, we selected Pangram v3 as the primary detector for our analyses, since it came out as the most consistent option. By contrast, Binoculars and DivEye proved to be rather unreliable. Desklib performed similarly in the text length and model family robustness analyses; it even outperformed Pangram v3 in model version robustness; however, it underperformed relative to Pangram v3 on HTML vs. plain text and language robustness, which we deem more important for this analysis. An additional advantage of Pangram v3 over the three other detectors is that it operates in a three-way classification scheme (AI-generated, AI-assisted, human) that provides richer signal than binary detection.

\FloatBarrier
\newpage

\section{Fact-Checking Annotation Interface} 
\label{sec:appendix_annotation}

\begin{figure}
    \centering
    \fbox{
        \includegraphics[width=\linewidth]{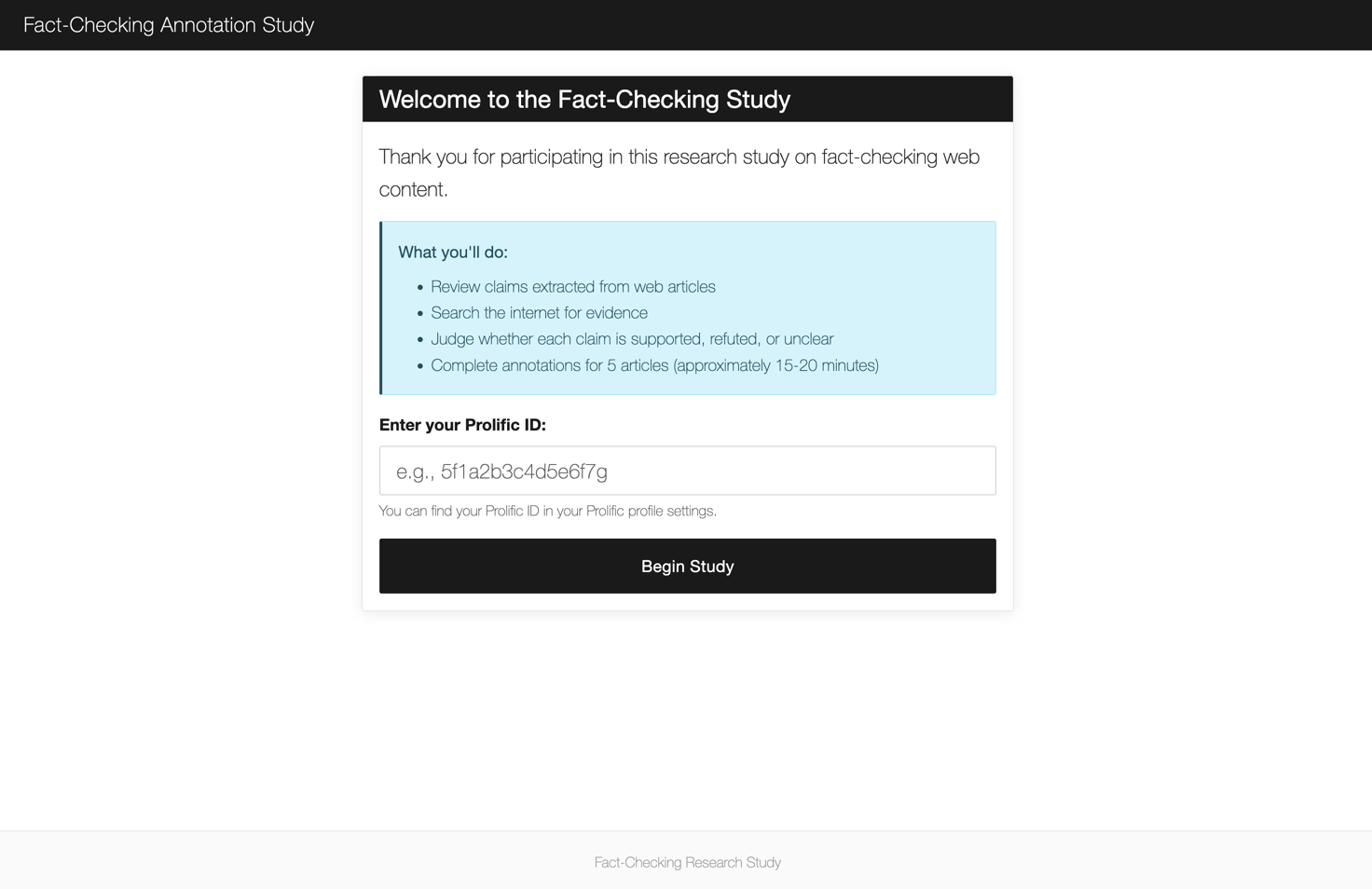}
    }
    \caption{\textbf{Welcome Screen.} The fact-checkers are presented with a list of expectations and asked to provide their Prolific IDs.}
    \label{fig:fc_app_0}
\end{figure}

The fact-checking annotation study (Hypothesis~2) was conducted using a custom web application built with Flask and deployed on DigitalOcean App Platform with a PostgreSQL database backend. Annotators were recruited through Prolific (see Section~\ref{sec:hypothesis_verification} for demographics and representativeness information) and accessed the application via a direct URL with their Prolific participant ID passed as a query parameter, as shown in Figure~\ref{fig:fc_app_0}.

\begin{figure}
    \centering
    \fbox{
        \includegraphics[width=\linewidth]{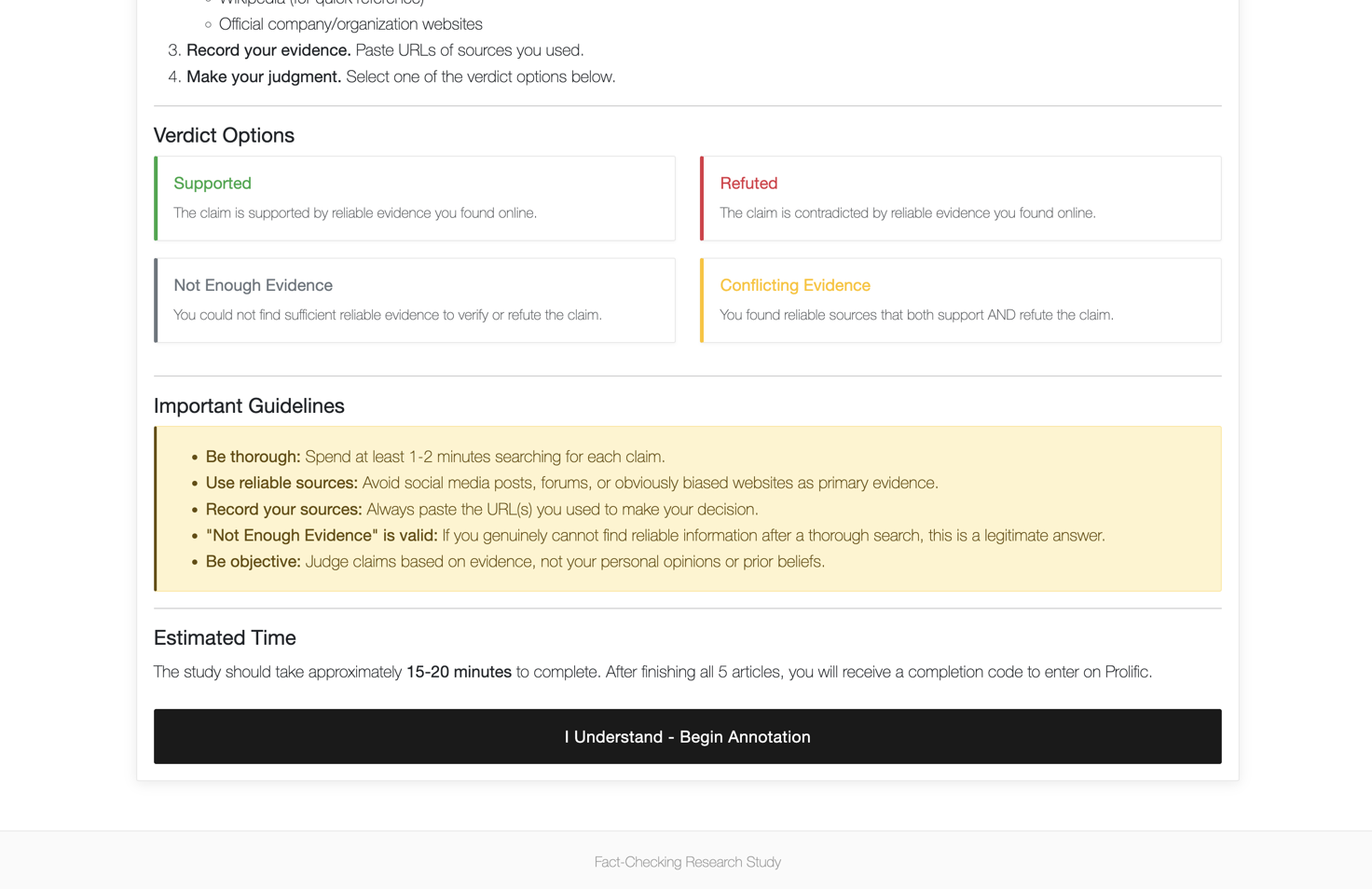}
    }
    \caption{\textbf{Instruction Recap Screens.} The fact-checkers are required to read through detailed instructions. The bottom of the screen is pictured.}
    \label{fig:fc_app_1}
\end{figure}

\noindent Upon entering the application, annotators were presented with an instructions page describing the task, the four verdict categories, and guidelines for evidence evaluation, whose ending is shown in Figure~\ref{fig:fc_app_1}. The instructions emphasized thoroughness (spending at least 1--2 minutes searching per claim), use of reliable sources, recording of evidence URLs, and objectivity. Annotators were told that ``Not Enough Evidence'' was a valid verdict when reliable information could not be found after a thorough search.

\begin{figure}
    \centering
    \fbox{
        \includegraphics[width=\linewidth]{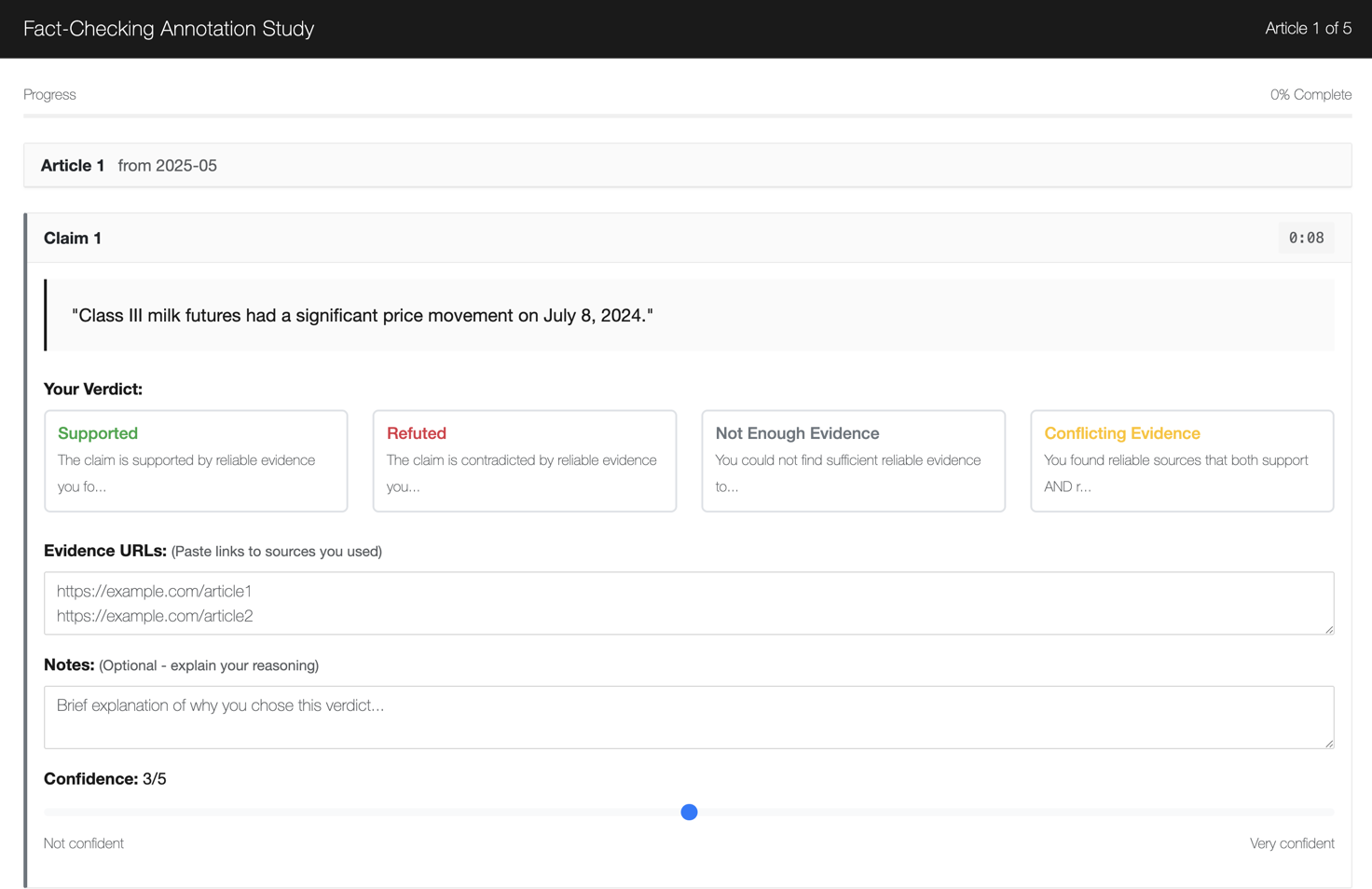}
    }
    \caption{\textbf{Item Annotation View.} Each claim for an article asks the fact-checkers to provide the verdict (Supported, Refuted, Not Enough Evidence, or Conflicting Evidence), evidence URLs, optional explanation, and confidence (range between 1, least confident, and 5, most confident). The top of the view indicates the progress within the assigned batch of articles.}
    \label{fig:fc_app_2}
\end{figure}

\begin{figure}
    \centering
    \fbox{
        \includegraphics[width=\linewidth]{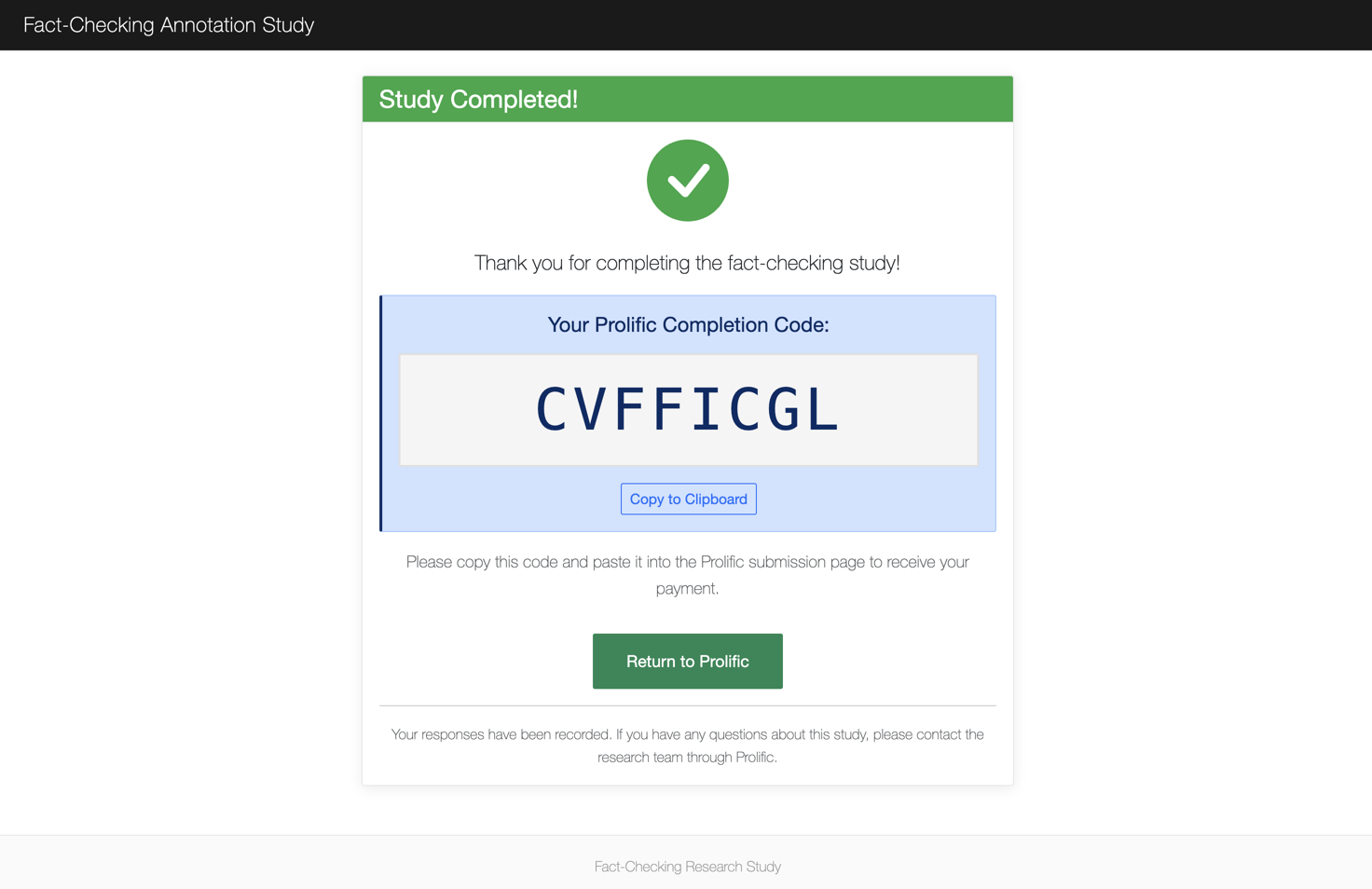}
    }
    \caption{\textbf{Completion Screen.} Upon completion, the fact-checkers are presented with the confirmation code for Prolific and instructions asking them to return to Prolific.}
    \label{fig:fc_app_3}
\end{figure}

\noindent The annotation interface presented claims grouped by source article. For each claim, the annotator was required to: (1)~select a verdict from the four-category scheme (Supported, Refuted, Not Enough Evidence, Conflicting Evidence), and (2)~provide a confidence rating on a 1--5 scale. Optionally, annotators could record evidence URLs and free-text notes. The annotation view for a single claim in displayed in Figure~\ref{fig:fc_app_2}. The interface tracked time spent per claim and per article. Upon completing all assigned articles, annotators received a completion code to submit on Prolific, as shown in Figure~\ref{fig:fc_app_3}.

\noindent The four-category verdict scheme was adapted from the FEVER~\citep{thorne2018fever} and AVeriTeC~\citep{schlichtkrull2024automated} annotation frameworks. Claim assignments were managed by the application to ensure approximately 20\% overlap across annotators, enabling computation of inter-annotator agreement via Krippendorff's alpha~\citep{krippendorff2004reliability}, Fleiss' kappa, and Cohen's kappa for pairwise comparisons.

\FloatBarrier
\newpage

\section{Additional Hypothesis Results}
\label{sec:appendix_additional_results}

This appendix presents the full set of quantitative analysis and participant study figures for Hypotheses~2, 4, 5, and~6, which are summarized in the main text (Section~\ref{sec:results}) but whose figures are omitted from the main body for space. For each hypothesis, we show (a)~the correlation between the measurable signal and the aggregate AI likelihood score across monthly samples, (b)~the overall distribution of participant survey responses, (c)~responses stratified by AI usage frequency, and (d)~responses stratified by general view of AI's impact on society.

\newpage

\subsection*{Hypothesis 2: Truth Decay}

\begin{figure}[h]
    \centering
    \begin{minipage}[t]{1.0\linewidth}
        \centering (a)
    \end{minipage}
    \includegraphics[width=1.0\linewidth]{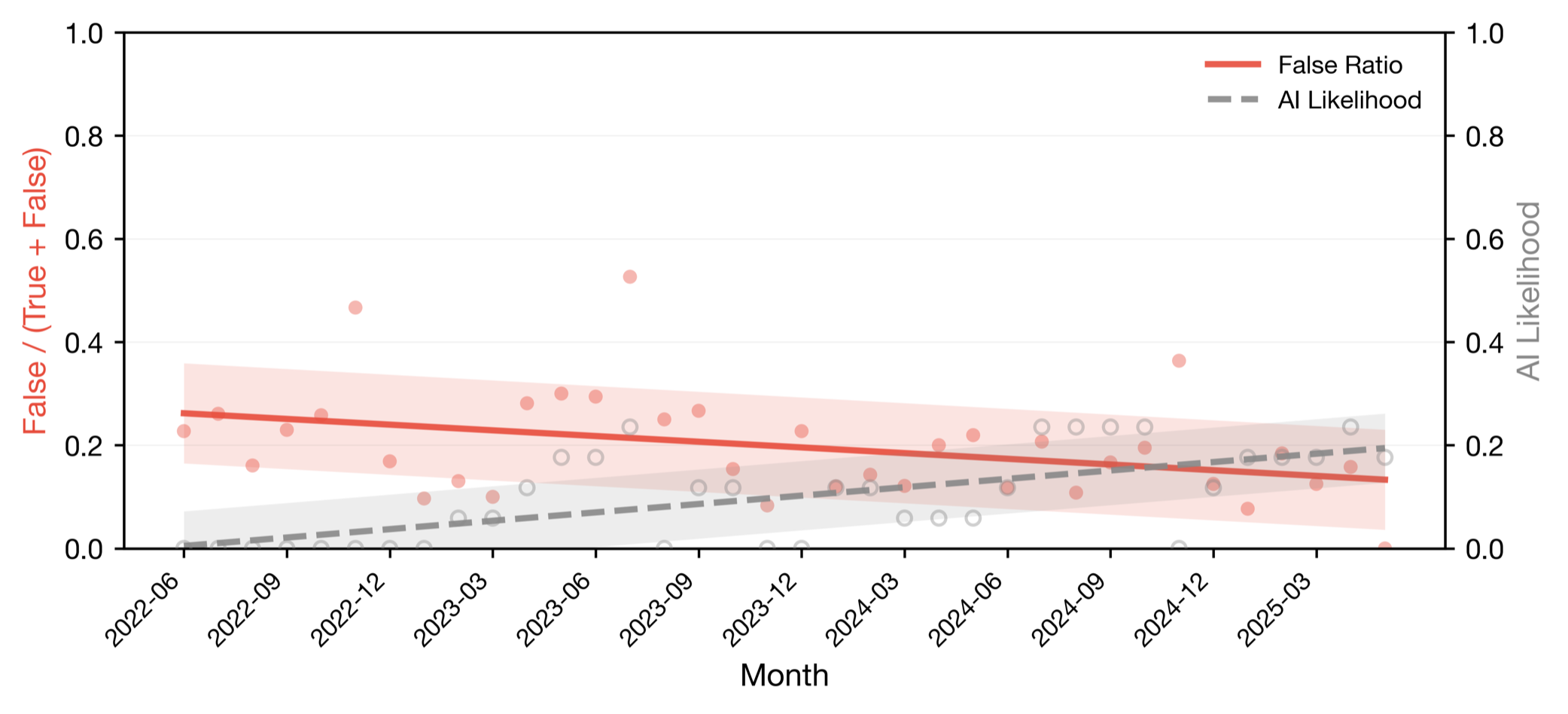}
    \vspace{1em}
    \includegraphics[width=1.0\linewidth]{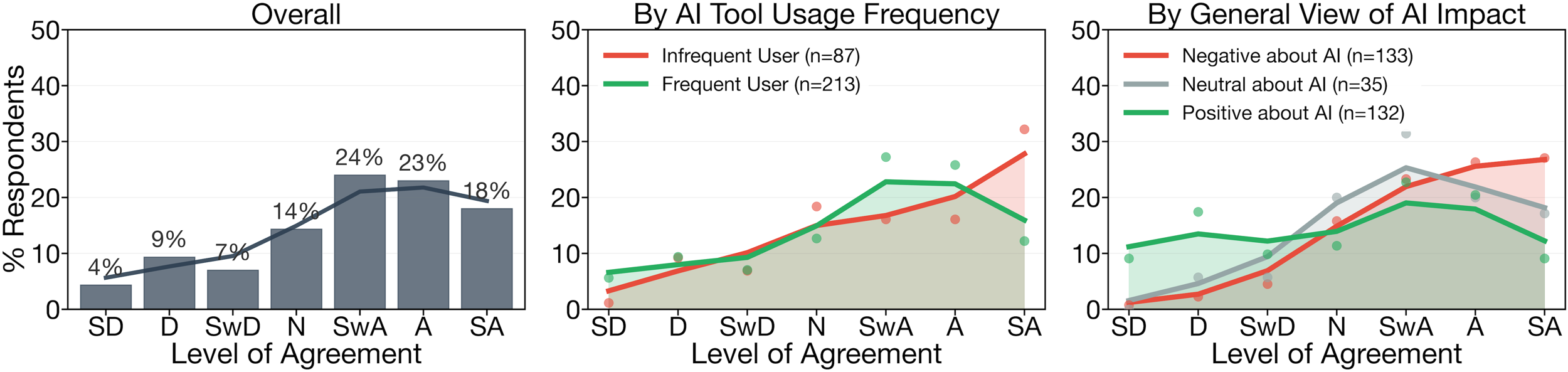}
    \vspace{0.5em}
    \begin{minipage}[t]{0.333\linewidth}
        \centering (b)
    \end{minipage}%
    \begin{minipage}[t]{0.333\linewidth}
        \centering (c)
    \end{minipage}%
    \begin{minipage}[t]{0.333\linewidth}
        \centering (d)
    \end{minipage}
    \caption{\textbf{Results for Hyp. 2: Truth Decay.} The figure shows results for the Truth Decay Hypothesis from the participant study (RQ1) and quantitative analysis of randomly sampled websites from the Internet Archive (RQ3). In (a), the average factual error rate is plotted against AI Likelihood score, as detected by Pangram v3 ($\rho=-0.19$, $p=0.27$). The overall results of the participant study are shown in (b), with responses ranging from Strongly Disagree (SD) to Strongly Agree (SA). These are broken down by AI usage frequency in (c) and general view of AI impact in (d).}
    \label{fig:appendix_hyp2}
\end{figure}

The Truth Decay Hypothesis posits that increasing AI-generated text on the internet leads to a higher rate of factually incorrect information. Figure~\ref{fig:appendix_hyp2} shows the quantitative analysis and participant study results. While $75.1\%$ of respondents leaned towards agreement with this hypothesis, our quantitative analysis did not find a statistically significant correlation between the factual error rate and the aggregate AI likelihood score ($\rho=-0.19$, $p=0.27$).

\FloatBarrier
\newpage

\subsection*{Hypothesis 4: Epistemic Islands}

\begin{figure}[h]
    \centering
    \begin{minipage}[t]{1.0\linewidth}
        \centering (a)
    \end{minipage}
    \includegraphics[width=1.0\linewidth]{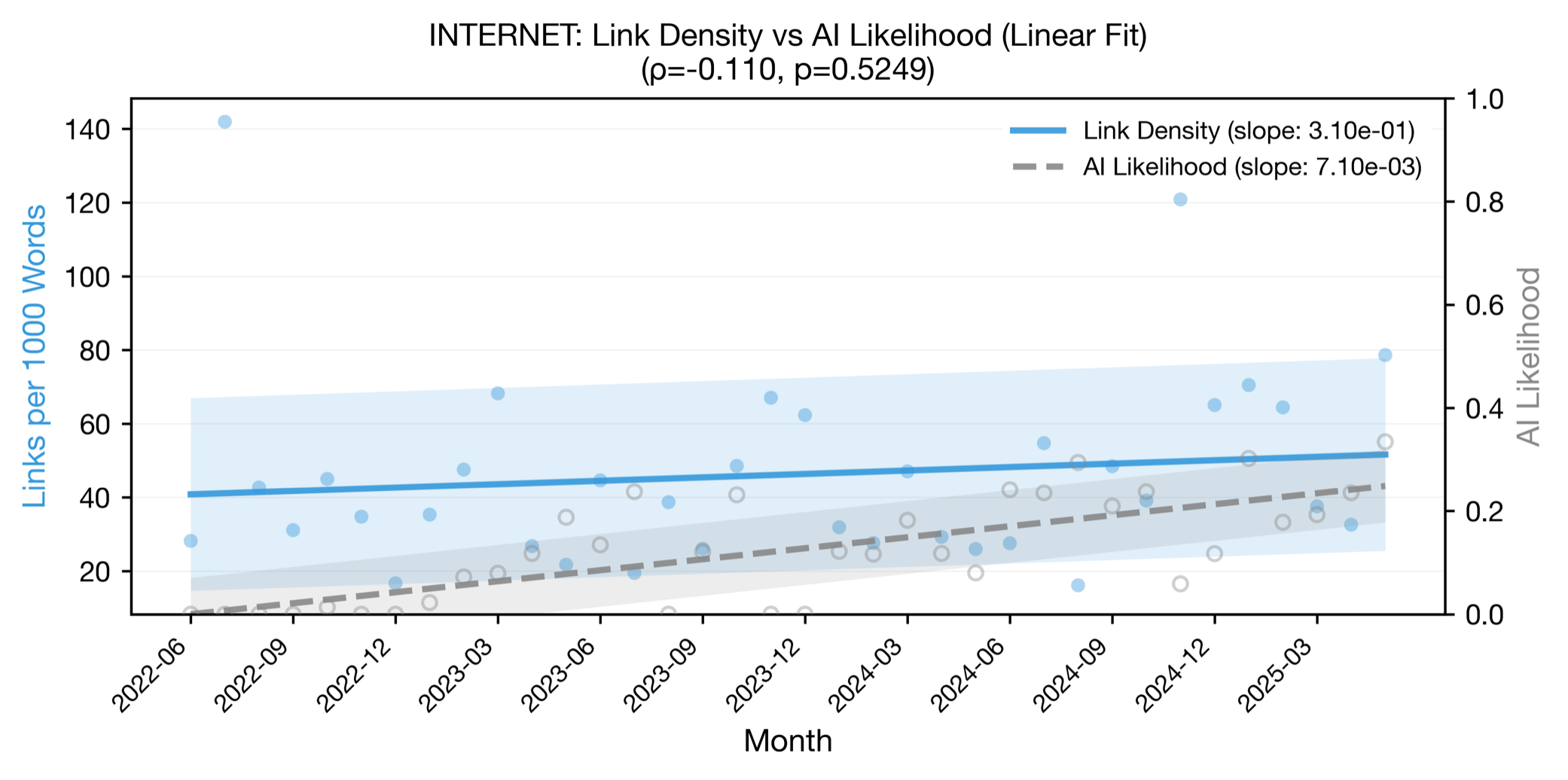}
    \vspace{1em}
    \includegraphics[width=1.0\linewidth]{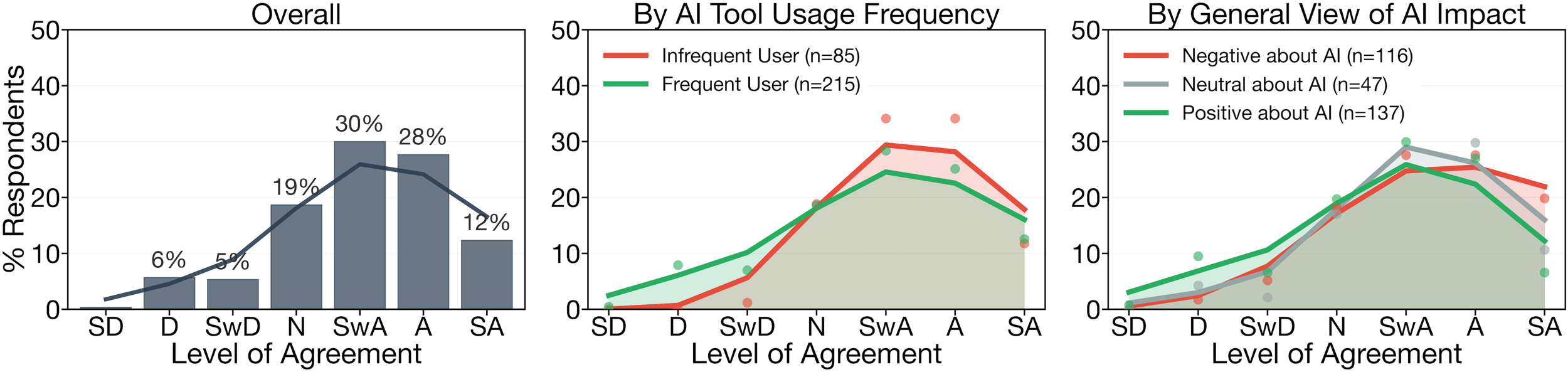}
    \vspace{0.5em}
    \begin{minipage}[t]{0.333\linewidth}
        \centering (b)
    \end{minipage}%
    \begin{minipage}[t]{0.333\linewidth}
        \centering (c)
    \end{minipage}%
    \begin{minipage}[t]{0.333\linewidth}
        \centering (d)
    \end{minipage}
    \caption{\textbf{Results for Hyp. 4: Epistemic Islands.} The figure shows results for the Epistemic Island Hypothesis from the participant study (RQ1) and quantitative analysis of randomly sampled websites from the Internet Archive (RQ3). In (a), the outbound link density (links per 1{,}000 words) is plotted against AI Likelihood score, as detected by Pangram v3 ($\rho=-0.12$, $p=0.48$). The overall results of the participant study are shown in (b), with responses ranging from Strongly Disagree (SD) to Strongly Agree (SA). These are broken down by AI usage frequency in (c) and general view of AI impact in (d).}
    \label{fig:appendix_hyp4}
\end{figure}

The Epistemic Island Hypothesis posits that as AI content becomes more common, articles increasingly provide answers without linking to external sources. Figure~\ref{fig:appendix_hyp4} shows the quantitative analysis and participant study results. While $69.9\%$ of respondents leaned towards agreement, the quantitative analysis did not find a statistically significant inverse correlation between link density and the aggregate AI likelihood score ($\rho=-0.12$, $p=0.48$).

\FloatBarrier
\newpage

\subsection*{Hypothesis 5: Entropy Dilution}

\begin{figure}[h]
    \centering
    \begin{minipage}[t]{1.0\linewidth}
        \centering (a)
    \end{minipage}
    \includegraphics[width=1.0\linewidth]{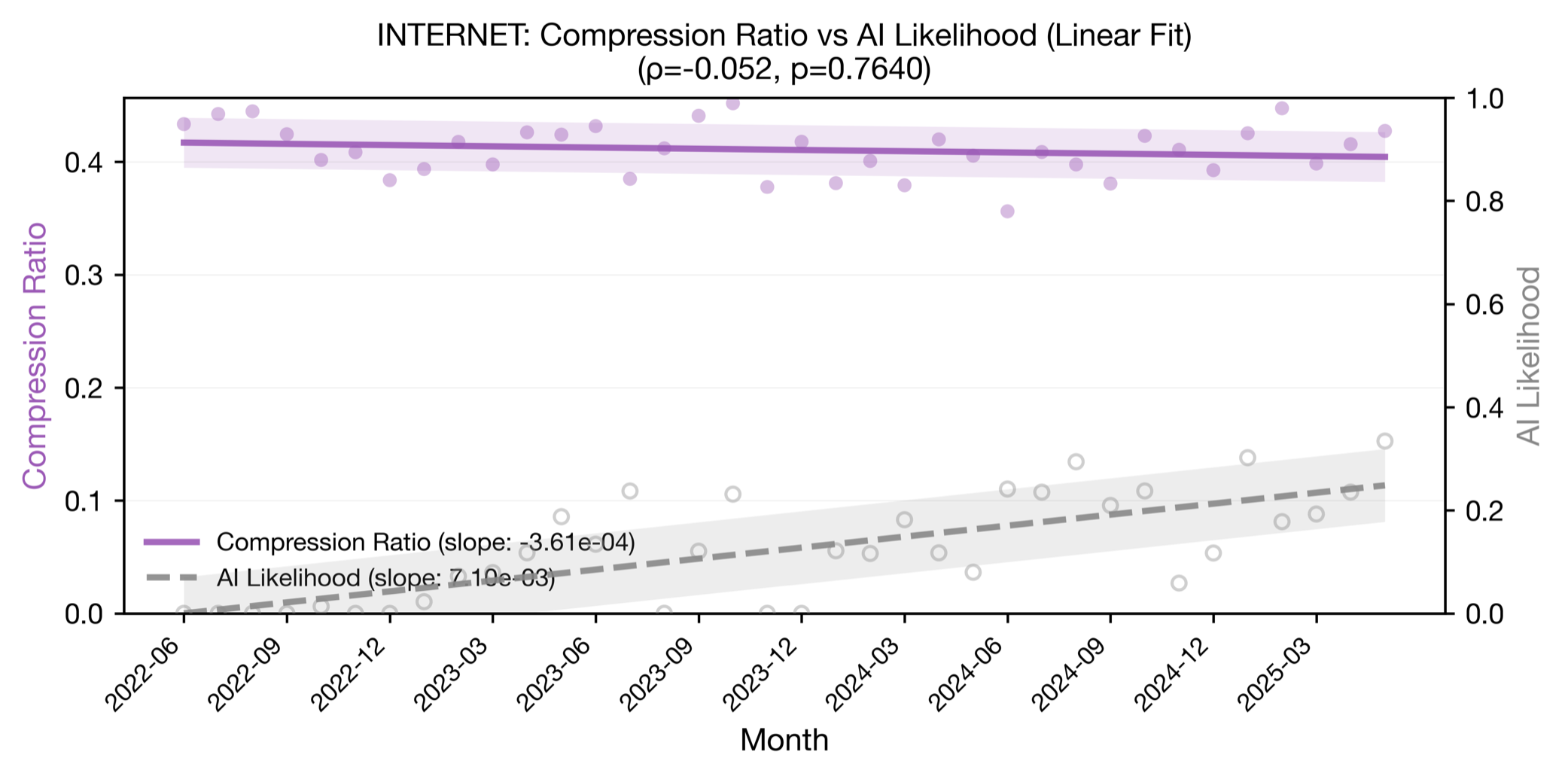}
    \vspace{1em}
    \includegraphics[width=1.0\linewidth]{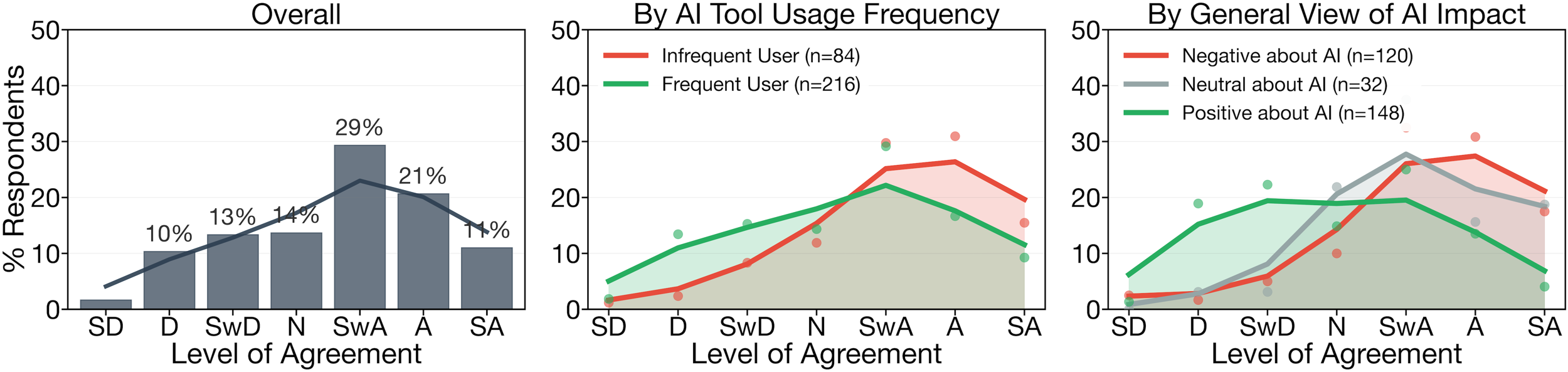}
    \vspace{0.5em}
    \begin{minipage}[t]{0.333\linewidth}
        \centering (b)
    \end{minipage}%
    \begin{minipage}[t]{0.333\linewidth}
        \centering (c)
    \end{minipage}%
    \begin{minipage}[t]{0.333\linewidth}
        \centering (d)
    \end{minipage}
    \caption{\textbf{Results for Hyp. 5: Entropy Dilution.} The figure shows results for the Entropy Dilution Hypothesis from the participant study (RQ1) and quantitative analysis of randomly sampled websites from the Internet Archive (RQ3). In (a), the Gzip compression ratio is plotted against AI Likelihood score, as detected by Pangram v3 ($\rho=-0.02$, $p=0.89$). The overall results of the participant study are shown in (b), with responses ranging from Strongly Disagree (SD) to Strongly Agree (SA). These are broken down by AI usage frequency in (c) and general view of AI impact in (d).}
    \label{fig:appendix_hyp5}
\end{figure}

The Entropy Dilution Hypothesis posits that as AI content becomes more common, content is becoming significantly longer while containing less actual meaning. Figure~\ref{fig:appendix_hyp5} shows the quantitative analysis and participant study results. While $60.7\%$ of respondents leaned towards agreement, the quantitative analysis did not find a statistically significant correlation between the Gzip compression ratio and the aggregate AI likelihood score ($\rho=-0.02$, $p=0.89$).

\FloatBarrier
\newpage

\subsection*{Hypothesis 6: Stylistic Monoculture}

\begin{figure}[h]
    \centering
    \begin{minipage}[t]{1.0\linewidth}
        \centering (a)
    \end{minipage}
    \includegraphics[width=1.0\linewidth]{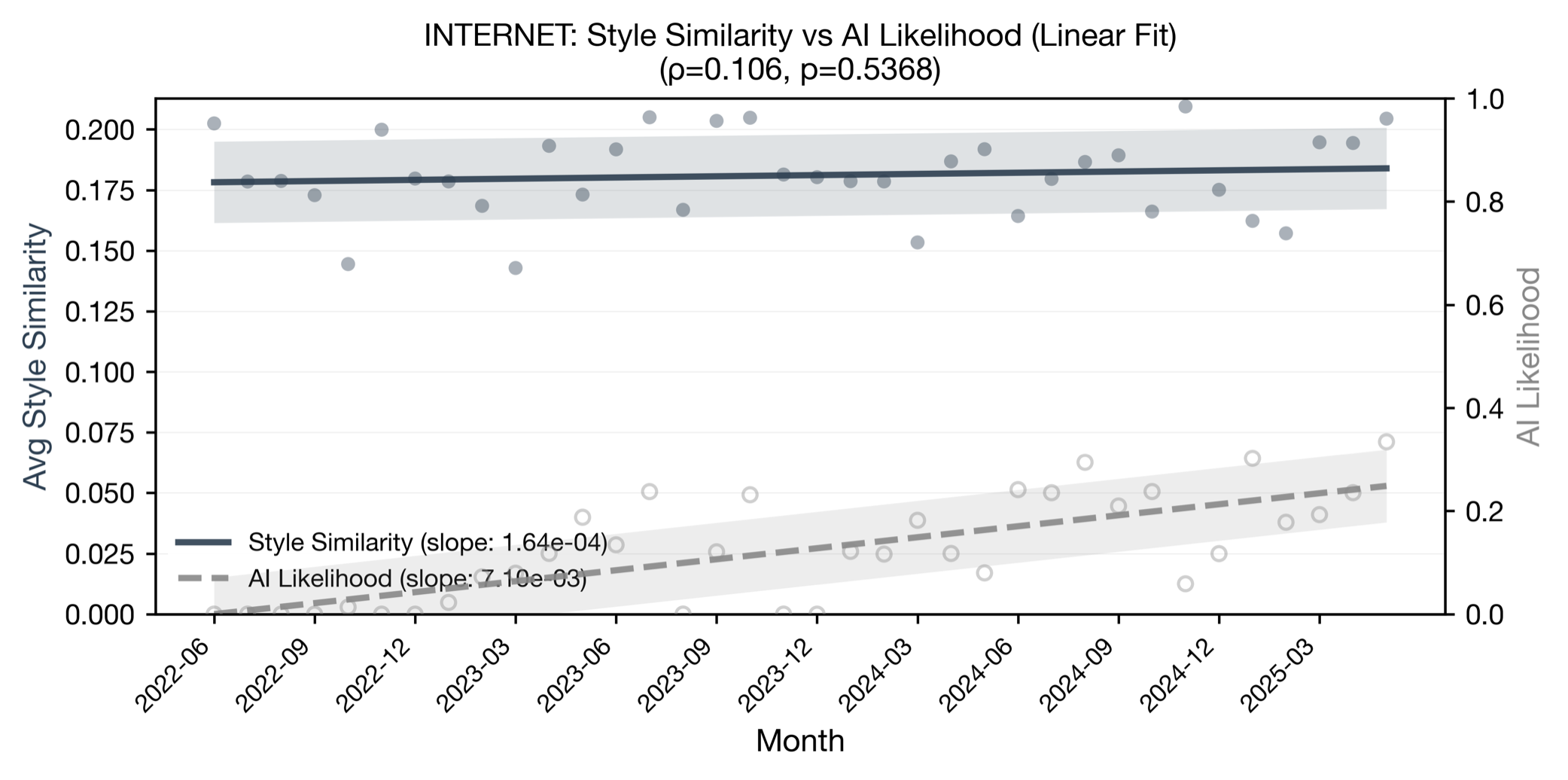}
    \vspace{1em}
    \includegraphics[width=1.0\linewidth]{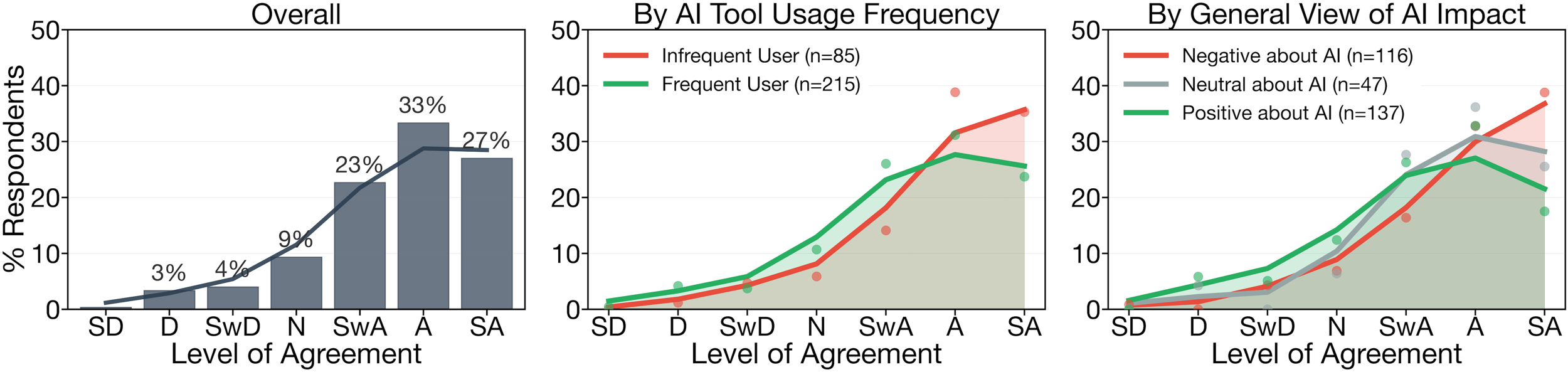}
    \vspace{0.5em}
    \begin{minipage}[t]{0.333\linewidth}
        \centering (b)
    \end{minipage}%
    \begin{minipage}[t]{0.333\linewidth}
        \centering (c)
    \end{minipage}%
    \begin{minipage}[t]{0.333\linewidth}
        \centering (d)
    \end{minipage}
    \caption{\textbf{Results for Hyp. 6: Stylistic Monoculture.} The figure shows results for the Stylistic Monoculture Hypothesis from the participant study (RQ1) and quantitative analysis of randomly sampled websites from the Internet Archive (RQ3). In (a), the average pairwise character 3-gram Jaccard similarity is plotted against AI Likelihood score, as detected by Pangram v3 ($\rho=0.24$, $p=0.17$). The overall results of the participant study are shown in (b), with responses ranging from Strongly Disagree (SD) to Strongly Agree (SA). These are broken down by AI usage frequency in (c) and general view of AI impact in (d).}
    \label{fig:appendix_hyp6}
\end{figure}

The Stylistic Monoculture Hypothesis posits that as AI content becomes more common, distinct individual writing styles are disappearing in favor of a generic, uniform voice. Figure~\ref{fig:appendix_hyp6} shows the quantitative analysis and participant study results. This hypothesis received the strongest agreement among participants ($83.0\%$ leaning towards agreement), yet the quantitative analysis did not find a statistically significant correlation between the average pairwise character 3-gram Jaccard similarity and the aggregate AI likelihood score ($\rho=0.24$, $p=0.17$).

\newpage
\section{Wayback Machine Sampling Details}
\label{sec:appendix_sampling}

Our sampling methodology follows~\citet{garg2025longitudinal}, with the following differences. First, we used a May 2025 ZipNum index and restricted our sample to URLs first archived on or after January 1, 2022, targeting the post-ChatGPT web rather than the complete web history. Second, we applied monthly temporal buckets with a target of $10{,}000$ URLs per month, as opposed to yearly buckets of $1$M URLs. Third, we applied a hard depth filter, excluding URLs with more than three path segments or more than two query parameters. Fourth, instead of logarithmic-scale downsampling, we applied a strict one-URL-per-host cap to prevent over-representation of frequently archived domains. Fifth, we explicitly required the first capture of each URL to return HTTP 200 with a \texttt{text/html} MIME type, verified via the CDX API. Finally, we shuffled each monthly bucket prior to sampling to neutralize the lexicographic ordering inherited from the ZipNum index. The last two steps were introduced because of the AI-generated text detection analysis.